\documentclass[a4paper,12pt]{iopart}
\usepackage{graphicx}
\usepackage{microtype}
\expandafter\let\csname equation*\endcsname\relax
\expandafter\let\csname endequation*\endcsname\relax
\usepackage{amsmath}
\usepackage{amssymb}
\usepackage{hyperref}
\hypersetup{pdfauthor={Darka Labavic, Hannes Nagel, Wolfhard Janke, Hildergard Meyer-Ortmanns},%
  pdftitle={Coarse-Grained Modeling of Genetic Circuits as a Function of the Inherent Time Scales},%
  hidelinks,allcolors=true}

\usepackage[disable]{todonotes}

\usepackage{float}
\usepackage{setspace}
\usepackage[english]{babel}
\usepackage{psfrag}
\usepackage{array}
\usepackage{color}

\begin{document}

\title[Coarse-Grained Modeling of Genetic Circuits as a Function of the Inherent \ldots]{Coarse-Grained Modeling of Genetic Circuits as a Function of the Inherent Time Scales}
\author{Darka Labavi\c{c}$^1$, Hannes Nagel$^2$, Wolfhard Janke$^2$ and Hildegard Meyer-Ortmanns$^1$}
\address{$^1$School of Engineering and Science, Jacobs University Bremen,\\ P.O.Box 750561, 28725 Bremen, Germany}
\address{$^2$Institut f\"ur Theoretische Physik, Universit\"at Leipzig,\\Postfach 100\,920, 04009 Leipzig, Germany}
\eads{\mailto{d.labavic@jacobs-university.de},  \mailto{hannes.nagel@itp.uni-leipzig.de}, \mailto{wolfhard.janke@itp.uni-leipzig.de} \mailto{h.ortmanns@jacobs-university.de}}

\begin{abstract}
  From a coarse-grained perspective the motif of a self-activating species, activating a second species which acts as its own repressor, is widely found in biological systems, in particular in genetic systems with inherent oscillatory behavior. Here we consider a specific realization of this motif as a genetic circuit, in which genes are described as directly producing proteins, leaving out the intermediate step of mRNA production. We focus on the effect that inherent time scales on the underlying fine-grained scale can have on the bifurcation patterns on a coarser scale in time. Time scales are set by the binding and unbinding rates of the transcription factors to the promoter regions of the genes. Depending on the ratio of these rates to the decay times of the proteins, the appropriate averaging procedure for obtaining a coarse-grained description changes and leads to sets of deterministic equations, which differ in their bifurcation structure. In particular the desired intermediate range of regular limit cycles fades away when the binding rates of genes are of the same order or less than the decay time of at least one of the proteins. Our analysis illustrates that the common topology of the widely found motif alone does not necessarily imply universal features in the dynamics.
\end{abstract}


\maketitle

\section{Introduction}\label{secI}
\label{intro_sec}
\begin{figure}[h]
  \begin{centering}
    \includegraphics[width=7cm]{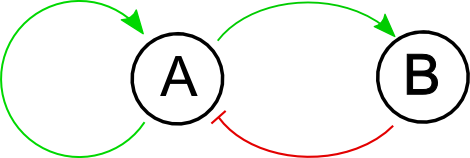}
    \par\end{centering}
  \caption{Basic motif of a self-activating species $A$, activating also its own repressor $B$. Pointed arrows denote activation, blunt arrow denotes repression.}\label{fig1}
\end{figure}

A frequently found motif in biological networks, in particular in genetic networks, is the combination of a positive feedback loop in which one species ($A$) activates itself, and a negative feedback loop, in which the first species activates its own repressor, the second species ($B$). In connection with genetic systems such motifs are realized in the cAMP signalling system in the slime mold Dictyosthelium Discoideum \cite{51}, in the embryonic division control system \cite{37,pomerening,52}, the MAPK-cascade \cite{mapk}, or in the circadian clock \cite{leibler,ingolia}. From the viewpoint of theoretical physics one is interested in dynamical features that are in common to the many different dynamical realizations of this motif. One common feature is certainly the occurrence of regular oscillations and the possibility of excitable behavior for an appropriate choice of parameters, and these features are captured already by a deterministic description in form of the bistable frustrated unit, considered in \cite{bartek} and references therein. It should be emphasized, however, that the different realizations do not only differ by the biological systems in which they are realized, but also by the degree in which this representation as two coupled loops as shown in Fig.~\ref{fig1}
 amounts only to an effective description. The effective description should be seen in contrast to a one-to-one mapping of the participating ingredients. In principle a number of intermediate steps may be included in these loops, and these intermediate steps need not be on the same level in case of a hierarchical organization. They could amount to reactions between genes leading to production rates on the level of proteins, or on the same level, just to reactions between proteins, and it may make a big difference if protein $A$ is repressed via the binding of a transcription factor of type $B$ to gene $b$ (design I), or via a direct repression of protein $A$ via $B$ (design II), as it was emphasized in \cite{guantes}. There it was shown that the distinct set of equations, corresponding to the two designs, lead to different oscillatory features and different behavior with respect to noise and external periodic signals, so that the motif in the first design acts as an ``integrator" of external stimuli, in the second design as a ``resonator".

\noindent For a coarse-grained description that should reflect the essential dynamical implications of this motif, the question therefore arises when it is actually justified to subsume the intermediate steps to effectively effective single steps and to represent the whole system by a set of two ordinary differential equations for the concentrations of $A$ and $B$. In particular such a description implies some kind of large-volume limit, in which the stochastic fluctuations of demographic origin, that is in the number of individuals, are ignored. In \cite{bartek} we have analyzed the effect of demographic fluctuations and of fluctuations in the reaction times in simple realizations of this motif, that is, without any intermediate steps. The stochastic fluctuations in the number of reactants and in the reaction times led to the occurrence of so-called quasi-cycles in parameter regimes for which the system would be deeply in the fixed-point regimes in the deterministic limit. On the other hand, the three regimes of the deterministic limit could be clearly recognized:
as function of one bifurcation parameter $\alpha$, the maximal rate of production of $A$ for full activation and no repression, the dynamics of the coupled species $A$ and $B$ converges to a fixed point and shows excitable behavior in a first phase. In a second phase, regular limit-cycle behavior is observed, and in a third phase, again a fixed point with excitable behavior is seen. %Both transitions between the phases are due to subcritical Hopf bifurcations.
The quasi-cycles which are only due to the finite system size, are more regular in the transition regions than far off in the fixed point phases, they can be distinguished from regular oscillations by the decay of their autocorrelation functions.

\noindent When this motif is supposed to describe a genetic circuit, a possible zoom into the dynamical details would amount to a description in terms of proteins $A$ and $B$, their associated mRNA at an intermediate level and two types of genes a and b, respectively. Protein production of type $A$ would result from transcription factors of type $A$, activating gene $a$  which is transcribed to the corresponding mRNA that leads to the translation to proteins $A$. At the same time, protein $A$ leads to an activation of gene $b$ via binding of the transcription factor $A$ to gene $b$, which is then transcribed  to mRNA of type $b$, and when translated to proteins leads to a repression of the production of $A$.
\vskip3pt
\noindent In general, such intermediate steps as the production of mRNA and the binding and unbinding of transcription factors to the promoter region of genes induce additional time scales: delay of the protein production, and the switching rate between gene states. Here we distinguish the following gene states: states with bound activating transcription factor (the gene being in the ``on-state"), states with bound repressing transcription factor (the gene being in the ``off-state"), or no transcription factor bound (which we call the gene being in the ``bare-state"). In principle we could consider further situations such as the simultaneous binding of an activator and a repressor to the promoter region of the gene $a$, but we leave out this option for simplicity. Whether the delay time and the switching rate can be ignored compared to the time scales defined by the protein decay rates of $A$ and $B$ is a matter of relative size. In addition, the excitable behavior with large excursions in phase space of the original motif is related to the pronounced difference in the decay times of the proteins: $A$ is considered as the fast variable and $B$ as the slow variable, similarly to the distinction of fast and slow variables in a FitzHugh-Nagumo element \cite{fitzhugh,nagumo}. (Due to this inherent difference, the oscillations become more spiky, the excursions in phase space from a fixed point more pronounced, and the description more realistic if we think of neurons realizing these circuits.) Therefore, when comparing the switching rate of gene states, we should specify with respect to which decay rate (that of $A$ or of $B$) it is fast or slow. In the following we call the switching rates ``fast" if they are fast in comparison to both decay rates, of $A$ and of $B$, ``slow" if the switching rate is of the order of the fast decay rate of $A$, and ``ultra-slow" if it is of the order of the slow decay rate of $B$.

\noindent In former related studies of this genetic circuit \cite{guantes} the switching rates of genes were assumed to be so high that their effect could be assumed to average out in the sense that mRNA and proteins see only average values of gene activation or repression. In this paper we want to analyze the effect that the inherent time scale of binding and unbinding rates of transcription factors has on the coarse-grained modeling. In order to project on this effect, we neglect the intermediate step of mRNA production, but vary the binding/unbinding rates to values which are no longer high, but of the same order as the decay times of either the ``fast" protein $A$ or of the ``slow" protein $B$.

\noindent From results for genetic switches, in particular for the toggle switch \cite{walczak}, a simpler system than ours with only two mutually repressing genes, it is known that additional fixed points may show up for (what we call) slow genes. More precisely, the joint probability distribution of the number of proteins in the cell and the DNA-binding site state (being on or off) have additional peaks, corresponding to additional fixed points in the deterministic limit, if the switches are studied in the nonadiabatic limit. If the switching rate between the genes is slow enough, the remainder of the system can follow the corresponding gene state of being either on or off. The dynamics of our system is more versatile. In addition to the fixed-point regime in the deterministic limit, we have the regime of regular limit-cycles, and the decay time of proteins differs by a factor of the order of hundred. So independently of the possibility to realize our system in synthetic genetic circuits or to find it in natural systems, we want to focus on the impact of ``slow" and ``ultra-slow" components on the system's dynamics, in particular on the phase structure.

\noindent We start from a fully stochastic description in terms of master equations describing reactions directly between genes and proteins, skipping the intermediate mRNA-production steps. We simulate these equations via the Gillespie algorithm \cite{gillespie}. In view of a coarse-grained description in the form of deterministic equations we take appropriate averages of the master equations which are adapted to the various limiting cases. These averaging procedures correspond to taking the zeroth order of a van Kampen expansion \cite{vankampen} of the master equations, where the expansion parameter is the inverse system size. The phase structure then differs for the resulting sets of deterministic equations, it depends on the competing time scales of genes and proteins. For fast genes we recover the previously observed structure of three regimes, two fixed-point and one limit-cycle regime. For this case we provide a detailed bifurcation analysis. For slow and ultra-slow genes, the regime of regular limit cycles is absent. Instead, the fast proteins $A$ see (certain combinations of) distinct states of gene expression separately rather than their averages: in the deterministic limit they therefore approach the fixed points corresponding to these states and stay there, unless the gene states change; in the stochastic realization they switch between these states, so that the probability density to find $N_A$ proteins of type $A$ and $N_B$ proteins of type $B$ shows as many peaks as there are gene states distinguished. In contrast, the inherent dynamics of the proteins $B$ is still so slow that these proteins cannot reach the vicinity of these fixed points in the stochastic realization which they would approach in the deterministic description, leading to broad peaks or almost uniform probability densities.

\noindent The subsequent sections of the paper are organized as follows. In sect.~\ref{secII} we present the model of interacting genes and proteins for the motif of a positive self-activating feedback loop coupled to a negative feedback loop with a second species repressing the first one. The model is described in terms of chemical reactions and master equations. In sect.~\ref{secIII} we derive deterministic models along with a stability analysis and a comparison with Gillespie simulations for three limiting cases: fast genes (sect.~\ref{secIII1}), slow genes (sect.~\ref{secIII2}) and ultra-slow genes (sect.~\ref{secIII3}). The summary and conclusions are given in sect.~\ref{secIV}, followed by an Appendix in which we analyze in detail the two transition regions between fixed-point behavior and limit cycles.

\section{The model}\label{secII}

\begin{figure}[b]
  \begin{centering}
    \includegraphics[width=10cm]{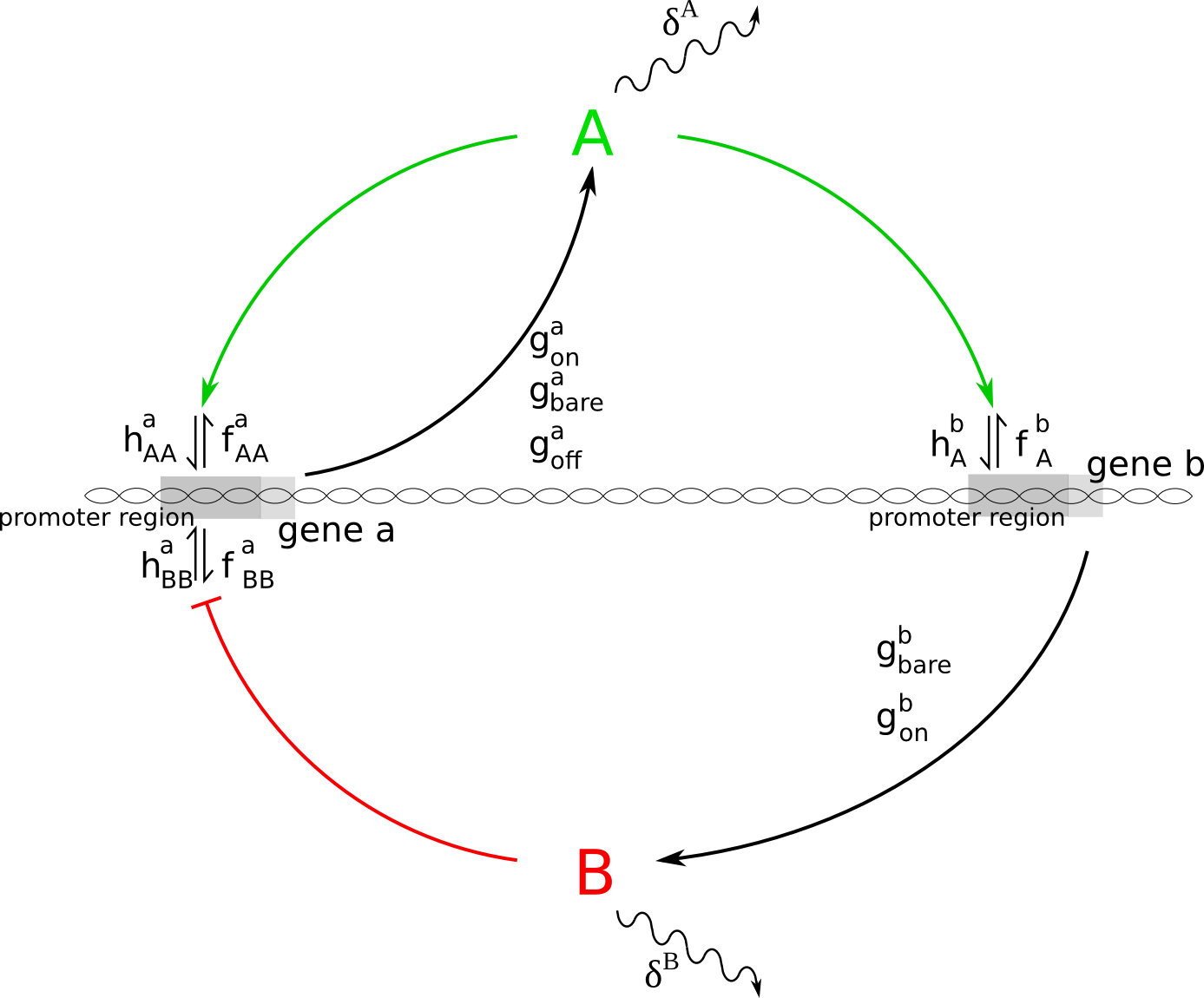}
    \caption{Zoom into the motif of Fig.~\ref{fig1} with a realization via genes $a$ and $b$ leading to the production of proteins $A$ and $B$ with rates $g^a_{\emph{on},\emph{bare},\emph{off}}$ and $g^b_{\emph{bare},\emph{on}}$, respectively, depending on the bound transcription factors to the promoter region of $a$ and $b$. Transcription factors $A(B)$ bind with rate $h^a_{AA}(h^a_{BB})$ to the promoter region of $a$ and unbind with rates $f^a_{AA}(f^a_{BB})$, respectively. Transcription factor $A$ also binds to the promoter region of $b$ with rate $h_A^a$ and unbinds with rate $f^b_A$. Proteins $A$ and $B$ decay with rates $\delta^A$ and $\delta^B$, respectively. For further explanations see the main text.}
    \par\end{centering}\label{fig2}
\end{figure}

In terms of biochemical reactions we consider the following realization of the motif of Fig.~\ref{fig1} displayed in Fig.~\ref{fig2}.

\noindent
Proteins $A$ and $B$ are produced under different conditions on the expression level of genes, but we assume in all cases that the production is proportional to the system size. The system size is parameterized by a factor $N_0$.  For protein $A$ we distinguish three situations: (i) An activating  transcription factor  $A$ is bound to the promoter region of gene $a$, so that gene $a$ is in the on-state and produces proteins $A$ accordingly with rate $g_{\emph{on}}^a\cdot N_0$; here the superscript $a$ and subscript $\emph{on}$ indicate that gene $a$ is responsible for the production of prbbotein $A$ and is itself in the on-state due to the binding of the (self-)activating transcription factor $A$. (ii) No transcription factor is bound to the promoter region of gene $a$, leading to a production of $A$ in the so-called bare state of the gene with rate $g^a_{\emph{bare}}\cdot N_0$. (iii) A repressing transcription factor $B$ is bound to the promoter region of gene $a$ and turns the gene to the lower expression level, so that we term the gene state to be ``off" and the production rate of protein $A$ proceeds with rate $g_{\emph{off}}^a\cdot N_0$, accordingly. For simplicity we leave out a further possible situation that an activating and a repressing transcription factor $A$ and $B$, respectively, are simultaneously bound to the respective promoter regions of gene $a$, leading to a conflicting input of activation and repression.

\noindent Since protein $B$ is only activated, but not repressed via $A$, we distinguish here only two states: (i) An activating transcription factor $A$ is bound to the promoter region of gene $b$, leading to a production of $B$ with rate $g^b_{\emph{on}}\cdot N_0$, or, (ii) no transcription factor is bound, leading to a production rate of $g^b_{\emph{bare}}\cdot N_0$ of protein $B$. Moreover, protein $A$ decays with rate $\delta^A$ and protein $B$ with rate $\delta^B$. Choosing $\delta^B$ much smaller than $\delta^A$ and $g^b_{\emph{on},\emph{bare}}$ much smaller than $g^a_{\emph{on}, \emph{bare}}$ is a way to implement the different inherent time scales in the protein dynamics: that of $A$ is much faster than that of $B$, the reason for why we call $A$ the fast variable and $B$ the slow variable also in this realization of the motif. The decay rate of the fast protein $A$ sets our time scale, $\delta^A=1$. Throughout this paper we choose $\delta^B=0.01$, so that the slow protein $B$ lives by a factor of 100 longer than $A$. The reactions referring to production and decay of proteins are summarized in the following equations:
\begin{equation}
  % \center
  \begin{array}{ccc}
    a_{\emph{on}}\; & \xrightarrow{\;\;g_{\emph{on}}^{a}N_0\;}             & \; A+a_{\emph{on}} \\
    \;\;\;a_{\emph{bare}}\;\;\;\;       & \xrightarrow{g_{\emph{bare}}^{a}N_0}           & \; A+a_{\emph{bare}} \\
    a_{\emph{off}}\;        & \xrightarrow{\;g_{\emph{off}}^{a}N_0\;}				& \; A+a_{\emph{off}} \\
    A\;              & \xrightarrow{\;\;\;\;\delta^{A}\;\;\;\;} 			& \;\phi\\
    b_{\emph{on}}\;         & \xrightarrow{\;\;g_{\emph{on}}^{b}N_0\;}				 & \; B+b_{\emph{on}}\\
    b_{\emph{bare}}\;       & \xrightarrow{g_{\emph{bare}}^{b}N_0}			 & \; B+b_{\emph{bare}} \\
    B\;              & \xrightarrow{\;\;\;\;\delta^{B}\;\;\;\;} 			& \;\phi\;.
  \end{array}
  \label{eq4}
\end{equation}

\noindent Next we come to the binding/unbinding reactions of transcription factors to the promoter region of genes. We distinguish between dimer binding of $A$ and $B$ to gene $a$ (corresponding to a Hill coefficient of $2$) and monomer binding of $A$ to the promoter region of gene $b$ (corresponding to a Hill coefficient of $1$). (The Hill coefficient provides a quantitative measure for the binding cooperativity.) The corresponding binding reactions along with the unbinding ones are listed in Eq.~(\ref{eq3}).
\begin{equation}
  % \center
  \begin{array}{ccc}
    a_{\emph{bare}}+2A\; & \xrightarrow{h_{AA}^{a}/N_0^2}              & \; a_{\emph{on}}\\
    a_{\emph{on}}\;      & \xrightarrow{\;\;\;f_{AA}^{a}\;\;\;\;}      & \; a_{\emph{bare}}+2A\\
    a_{\emph{bare}}+2B\; & \xrightarrow{h_{BB}^{a}/N_0^2}              & \; a_{\emph{off}}\\
    a_{\emph{off}}\;     & \xrightarrow{\;\;\;f_{BB}^{a}\;\;\;\;}      & \; a_{\emph{bare}}+2B\\
    b_{\emph{bare}}+A\;  & \xrightarrow{\;\;h_{A}^{b}/N_0\;}           & \; b_{\emph{on}}\\
    b_{\emph{on}}\;      & \xrightarrow{\;\;\;\;\;f_{A}^{b}\;\;\;\;\;} & \; b_{\emph{bare}}+A\;.\\
  \end{array}\label{eq3}
\end{equation}

\noindent The notation is chosen as follows: The superscript in the binding/unbinding coefficients $h^i$, $f^j$ indicate the gene whose promoter region is affected in the binding/unbinding event, the subscripts refer to the monomer (one index) or dimer (two indices) binding/unbinding of the transcription factors. In our Gillespie simulations we choose the effective binding rates and the corresponding unbinding rates to be of the same order such that
\begin{equation}
  \frac{h^a_{AA}N_A^2}{N_0^2}=\frac{h^a_{BB}N_B^2}{N_0^2}\;=\;\frac{h_A^b N_A}{N_0}
  \sim f^a_{AA}\;=\;f^a_{BB}\;=\;f^b_A\;.
\end{equation}
We normalize the binding rates which are proportional to the number of proteins $N_A$ with the appropriate power in $N_0$ \todo{Hill Coefficient} to make them approximately independent of the system size, assuming that $N_0\sim N_A$. Our simulations have shown that in case of a ratio of binding and unbinding rates different from one, it is the smaller value of the binding and unbinding rates that determines the dynamics in the following sense: The implied changes affect only the location of the fixed points and the limit-cycle regime, but lead to no qualitative change.

\noindent Given now the common values for the binding and unbinding rates of transcription factors, we call the switching of gene states, induced by the binding and unbinding events, fast if these rates are much higher than the decay rate of the fast protein (set to one), slow if it is of the order of the fast protein~$A$, and ultra-slow if it is of the order of the slow protein~$B$.

%%% 
%%% TODO (removed parts)
%%% 

\noindent As we shall argue in sect.~\ref{secIII1}, the $A$-protein production rate in the on-state, $g^a_{\emph{on}}$, is chosen as the bifurcation parameter, while the production rates in the other states of gene $a$ are kept fixed, such that $g^a_{\emph{bare}}$ is by an order of magnitude smaller than the usual values of $g^a_{\emph{on}}$, and $g^a_{\emph{off}}$ is set to zero. The production rates of the protein $B$ are also kept fixed and chosen by two orders of magnitude smaller than the corresponding production rates of protein $A$ to compensate for the hundred times longer lifetime of protein $B$ in the competing gain and loss terms in Eq.~(\ref{eqdetfastgenes}), see Sect.~\ref{secIII1}. Our choice of parameters is summarized in Tables~\ref{table1} and \ref{table2}.

\begin{table}[ht]
  % \begin{center}
  \caption{Fixed parameters.}\label{table1}
  % \centering{}%
  \begin{tabular}{|c|c|c|c|c|c|}
    \hline
    $g^a_{\emph{bare}}$ & $g^a_{\emph{off}}$ &$g^b_{\emph{on}}$& $g^b_{\emph{bare}}$ & $\delta^{A}$&$\delta^{B}$\tabularnewline
    \hline
    25 & 0 & 2.5& 0.025 & 1&0.01\tabularnewline
    \hline
  \end{tabular}
  % \par\end{centering}
  % \end{center}
\end{table}

\begin{table}[h]
  \begin{centering}
    \caption{Binding and unbinding parameters.}\label{table2}
    \centering{}%
    \begin{tabular}{|c|c|c|c|c|c|c|c|}
      \hline
      genes&$N_0$&$N_A$&$N_B$& $f_{AA}^{a}=f_{BB}^{a}=f_{A}^{b}$ & $\frac{h_{AA}^{a}}{N_0^2}N_A^2=\frac{h_{BB}^{a}}{N_0^2}N_B^2$ & $\frac{h_{A}^{b}}{N_0}N_A$ &  $\delta^{A}\gg\delta^{B}$\tabularnewline
      \hline
      \hline
      fast&1&100&100 & 100 &100 & 100 & $\gg\delta^{A}\gg\delta^{B}$\tabularnewline
      \hline
      slow &1&100&100 & 1 & 1 & 1 & $\sim\delta^{A}\gg\delta^{B}$\tabularnewline
      \hline
      ultra-slow&1&100&100 & 0.01 & 0.01 & 0.01 & $\sim\delta^{B}\ll\delta^{A}$\tabularnewline
      \hline
    \end{tabular}
    \par\end{centering}
\end{table}

\noindent
The reactions described by Eqs.(\ref{eq4}) and (\ref{eq3}) correspond to a set of master equations which tell us the change in time of the probability to find at time $t$ $N_A$ proteins of type $A$ and $N_B$ proteins of type $B$, given that gene $a$ is in either of the three states $i=\emph{on}, \emph{bare}, \emph{off}$ and at the same time gene $b$ is either in the on-state $j=\emph{on}$, or in the bare-state $j=\emph{bare}$. This probability is denoted as $P_{ij}(N_A,N_B,t)$. The six master equations for $P_{ij}(N_A,N_B,t)$, resulting from the six combinations of indices, can be summarized in matrix notation according to:
\begin{eqnarray}\label{eqmaster}
  \frac{dP_{i,j}(N_{A},N_{B};t)}{dt} & = & -\left(g_{i}^{a}N_{0}+\delta^{A}N_{A}+g_{j}^{b}N_{0}+\delta^{B}N_{B}\right)P_{i,j}(N_{A},N_{B},t)\nonumber \\
  & + & g_{i}^{a}N_{0}P_{i,j}(N_{A}-1,N_{B},t)+\delta^{A}(N_{A}+1)P_{i,j}(N_{A}+1,N_{B},t)\nonumber \\
  & + & g_{j}^{b}N_{0}P_{i,j}(N_{A},N_{B}-1,t)+\delta^{B}(N_{B}+1)P_{i,j}(N_{A},N_{B}+1,t)\nonumber\\
  & + & h_{AA}^{a}(i)\frac{N_{A}^{2}}{N_{0}^{2}}P_{\emph{bare},j}(N_{A},N_{B},t)-f_{AA}^{a}(i)P_{\emph{on},j}(N_{A},N_{B},t)\nonumber \\
  & + & h_{BB}^{a}(i)\frac{N_{B}^2}{N_{0}^2}P_{\emph{bare},j}(N_{A},N_{B},t)-f_{BB}^{a}(i)P_{\emph{off},j}(N_{A},N_{B},t)\nonumber \\
  & + & h_{A}^{b}(j)\frac{N_{A}}{N_{0}}P_{i,\emph{bare}}(N_{A},N_{B},t)-f_{A}^{b}(j)P_{i,\emph{on}}(N_{A},N_{B},t)\nonumber\\
  &i&= \emph{on},\,\emph{bare},\,\emph{off},\; j=\emph{on},\,\emph{bare},
\end{eqnarray}
if we introduce the following definitions:
\begin{eqnarray}
  h_{AA}^{a}(\emph{on})&=&-h_{AA}^{a}(\emph{bare})\equiv h_{AA}^{a},\nonumber\\
  h_{BB}^{a}(\emph{off})&=&-h_{BB}^a(\emph{bare})\equiv h_{BB}^{a},\nonumber\\
  h_{AA}^{a}(\emph{off})&=&\;\; h_{BB}^{a}(\emph{on})=0
\end{eqnarray}
for the dimer binding factors, and
\begin{eqnarray}
  f_{AA}^{a}(\emph{on})&=&-f_{AA}^{a}(\emph{bare})\equiv f_{AA}^{a},\nonumber\\
  f_{BB}^{a}(\emph{off})&=&-f_{BB}^a(\emph{bare})\equiv f_{BB}^{a},\nonumber\\
  f_{AA}^{a}(\emph{off})&=&\;\;f_{BB}^{a}(\emph{on})=0
\end{eqnarray}
for the dimer unbinding, where the argument in the bracket refers to the state of either gene $a$ or gene $b$, referred to via $i$ or $j$ in the master equations, respectively. For monomer binding/unbinding factors we  define
\begin{eqnarray}
  h_A^b(\emph{on})&=&-h_A^b(\emph{bare})\equiv h_A^b, \nonumber\\
  f_A^b(\emph{on})&=&-f_A^b(\emph{bare})\equiv f_A^b\;.
\end{eqnarray}
The first eight terms of the master equation for $P_{ij}(N_A,N_B,t)$ for all values of $i,j$ result from the production or decay of proteins, leading to gain or loss terms as follows: Loss terms contributing to the change of $P_{ij}(N_A,N_B,t)$ are due to the production of protein $A$ with constant rate $g^a_i\cdot N_0$, or the decay of $A$ proportional to $N_A$, and the production of protein $B$  with rate $g^b_j\cdot N_0$, or the decay of $B$ proportional to $N_B$. Gain terms, on the other hand, are due to the production of $A$ from a state with $N_A-1$ proteins $A$, and the production of $B$ from a state with $N_B-1$ proteins $B$, or the decay of one protein $A$ from a state with $N_A+1$ proteins $A$, or the decay of one protein $B$ from a state with $N_B+1$ proteins $B$.

\noindent The last six terms in Eq.~(\ref{eqmaster}), of which between four and six are different from zero, describe changes in the probability due to the binding and unbinding of proteins to the promoter regions of gene $a$ and $b$. For example, a positive contribution to the probability $P_{\emph{on},j}(N_A,N_B,t)$ results from a binding of a dimer of proteins $A$ with rate $h^a_{AA}(\emph{on})\frac{N_A^2}{N_0^2}$, given that the system before the binding event has $N_A,N_B$ proteins with gene $a$ being in the bare state; a negative contribution results from an unbinding of a dimer of $A$-proteins with rate $f_{AA}^a(\emph{on})=f^a_{AA}$, given that the system contains $N_A,N_B$ proteins while gene $a$ being in the on-state before the unbinding event. The other terms are derived similarly.

\section{Coarse-grained description of the genetic circuit}\label{secIII}
The coarse-graining that we intend to achieve refers primarily to a coarse-graining in time, averaging over events on a time scale much shorter than the scale on which the effective description should hold. (Usually, along with the coarse-graining in time goes a coarse-graining in space, as it is familiar from other realms of physics, but this is not in our focus here, as we do not arrange the genetic circuit and its constituents in any spatial ordering. Here the coarse-graining in time leads in general to a reduction of variables that are needed to describe the system's dynamics, but the amount of reduction depends on the limits which are taken, as we shall see below.) In order to derive coarse-grained descriptions in the form of deterministic equations for the protein concentrations $\Phi_A:=N_A/N_0$ and $\Phi_B:=N_B/N_0$, we apply appropriate averaging procedures to the master equation (\ref{eqmaster}). We assume that the states of the genes are independent of each other and of the number of proteins $N_A$ and $N_B$. Therefore we factorize the probability $P_{ij}(N_A,N_B,t)$ according to
\begin{equation}
  \label{eqfac}
  P_{ij}(N_A,N_B,t)\;=\;a_ib_j\;P(N_A,N_B,t),
\end{equation}
with $a_i$ the probability of finding gene $a$ in the $i$-state and $b_j$ the probability of finding gene $b$ in the $j$-state, while $P(N_A,N_B,t)$ is the probability of finding the respective protein numbers whatever states the genes are in. We then consider expectation values
\begin{eqnarray}\label{eqns}
  \langle N_S\rangle_{ij}&\equiv&\sum_{N_A,N_B} P_{ij}(N_A,N_B,t)\;N_S\nonumber\\
  &=&\sum_{N_A,N_B} a_i b_j P(N_A,N_B,t)\;N_S\nonumber\\
  &=& \langle N_S\rangle a_i b_j\;,
\end{eqnarray}
where $S$ stands for the species $A$ or $B$ and the average represents a summation over all $N_A$, $N_B$ values. We have to postulate
\begin{equation}
  \sum_i a_i\;=\;1\;=\;\sum_j b_j\;=\sum_{N_A,N_B} P(N_A,N_B,t),
\end{equation}
so that
\begin{equation}\label{eqaibj}
  a_i\;=\;\sum_{N_A,N_B,j}P_{ij}(N_A,N_B,t),\quad b_j\;=\sum_{N_A,N_B,i}P_{ij}(N_A,N_B,t)
\end{equation}
and
\begin{equation}\label{eq8}
  \sum_{N_{A}N_{B}}\frac{dP_{ij}(N_{A},N_{B})}{dt}=\frac{d(a_{i}b_{j})}{dt},\;\sum_{j}\frac{d(a_{i}b_{j})}{dt}
  =\frac{da_{i}}{dt},\;\sum_{i}\frac{d(a_{i}b_{j})}{dt}=\frac{db_{j}}{dt}.
\end{equation}
Moreover, using (\ref{eq8}) and (\ref{eqns}), we have
\begin{equation}
  \sum_{N_{A}N_{B}}\frac{dP_{ij}(N_{A},N_{B},t)}{dt}N_{S}=\frac{d(\langle N_{S}\rangle a_{i}b_{j})}{dt}\;;\qquad
  \sum_{i,j}\langle N_{S}\rangle a_{i}b_{j}=\langle N_{S}\rangle
\end{equation}
and
\begin{equation}
  \sum_{i,j}\frac{d(\langle N_{S}\rangle a_{i}b_{j}}{dt}=\frac{d\langle N_{S}\rangle }{dt}.
\end{equation}
Now we multiply the master equation (\ref{eqmaster}) with $N_A$ and $N_B$, respectively, and sum over all $N_A,N_B$; next we sum the average value of $N_A$, $\langle N_A\rangle$ only over all states of gene $b$, since the change in $N_A$ is assumed to be independent on the states of gene $b$, and in analogy $\langle N_B\rangle$ only over all states of gene $a$, respectively. We then obtain
\begin{equation}
  \frac{d}{dt}\left(\sum_{N_A,N_B,j}N_AP_{ij}(N_A,N_B,t)\right)\;=\;\frac{d}{dt}(\langle N_A\rangle a_i),
\end{equation}
where
\begin{eqnarray}
  \frac{d\left(\langle N_{A}\rangle a_{\emph{on}}\right)}{dt}&=&g_{\emph{on}}^{a}N_{0}a_{\emph{on}}+h_{AA}^{a}\frac{\langle N_{A}\rangle ^{3}}{N_{0}^{2}}a_{\emph{bare}}-f_{AA}^{a}\langle N_{A}\rangle a_{\emph{on}}-\delta^{A}\langle N_{A}\rangle a_{\emph{on}}\label{eq13}\\
  \frac{d(\langle N_{A}\rangle a_{\emph{off}})}{dt}&=&g_{\emph{off}}^{a}N_{0}a_{\emph{off}}+h_{BB}^{a}\langle N_{A}\rangle \frac{\langle N_{B}\rangle^2 }{N_{0}^2}a_{\emph{bare}}-f_{BB}^{a}\langle N_{A}\rangle a_{\emph{off}}\nonumber\\
  &&-\delta^{A}\langle N_{A}\rangle a_{\emph{off}}\label{eq14}\\
  \frac{d\left(\langle N_{A}\rangle a_{\emph{bare}}\right)}{dt} &=& g_{\emph{bare}}^{a}N_{0}a_{\emph{bare}}-\delta^{A}\langle N_{A}\rangle a_{\emph{bare}} -h_{AA}^{a}\frac{\langle N_{A}\rangle ^{3}}{N_{0}^{2}}a_{\emph{bare}}
  +f_{AA}^{a}\langle N_{A}\rangle a_{\emph{on}}\nonumber\\
  &&-h_{BB}^{a}\langle N_{A}\rangle \frac{\langle N_{B}\rangle^2 }{N_{0}^2}a_{\emph{bare}}+f_{BB}^{a}\langle N_{A}\rangle a_{\emph{off}}\;.\label{eq15}
\end{eqnarray}
Similarly,
\begin{equation}
  \frac{d}{dt}\left(\sum_{N_A,N_B,i}N_BP_{ij}(N_A,N_B,t)\right)\;=\;\frac{d}{dt}(\langle N_B\rangle b_j),
\end{equation}
where
\begin{eqnarray}
  \frac{d(\langle N_{B}\rangle b_{\emph{on}})}{dt}&=&g_{\emph{on}}^{b}N_{0}b_{\emph{on}}-\delta^{B}\langle N_{B}\rangle b_{\emph{on}}+h_{A}^{b}\frac{\langle N_{A}\rangle }{N_{0}}\langle N_{B}\rangle b_{\emph{bare}}-f_{A}^{b}\langle N_{B}\rangle b_{\emph{on}}\label{eq16}\\
  \frac{d(\langle N_{B}\rangle b_{\emph{bare}})}{dt}&=&g_{\emph{bare}}^{b}N_{0}b_{\emph{bare}}-\delta^{B}\langle N_{B}\rangle b_{\emph{bare}}-h_{A}^{b}\frac{\langle N_{A}\rangle }{N_{0}}\langle N_{B}\rangle b_{\emph{bare}}+f_{A}^{b}\langle N_{B}\rangle b_{\emph{on}}.\nonumber\\
  {}\label{eq17}
\end{eqnarray}
{}From Eq.~(\ref{eqaibj}) we obtain
\begin{eqnarray}
  \frac{da_{\emph{on}}}{dt} & = & h_{AA}^{a}\frac{\left\langle N_{A}\right\rangle ^{2}}{N_{0}}a_{\emph{bare}}-f_{AA}^{a}a_{\emph{on}}\nonumber \\
  \frac{da_{\emph{bare}}}{dt} & = & -h_{AA}^{a}\frac{\left\langle N_{A}\right\rangle ^{2}}{N_{0}}a_{\emph{bare}}+f_{AA}^{a}a_{\emph{on}}-h_{BB}^{a}\frac{\left\langle N_{B}^2\right\rangle }{N_{0}^2}a_{\emph{bare}}+f_{BB}^{a}a_{\emph{off}}\nonumber \\
  \frac{da_{\emph{off}}}{dt} & = & h_{BB}^{a}\frac{\left\langle N_{B}\right\rangle^2 }{N_{0}^2}a_{\emph{bare}}-f_{BB}^{a}a_{\emph{off}}\nonumber \\
  \frac{db_{\emph{on}}}{dt} & = & h_{A}^{b}\frac{\left\langle N_{A}\right\rangle }{N_{0}}b_{\emph{bare}}-f_{A}^{b}b_{\emph{on}}\nonumber \\
  \frac{db_{\emph{bare}}}{dt} & = & -h_{A}^{b}\frac{\left\langle N_{A}\right\rangle }{N_{0}}b_{\emph{bare}}+f_{A}^{b}b_{\emph{on}}.\label{eq18}
\end{eqnarray}
These equations determine the time dependence of the probability to find the system in any of the five different gene states. In a stationary state of the genes, the left-hand side of Eqs.~(\ref{eq18}) vanishes. This leads to
\begin{eqnarray}
  a_{\emph{on}}&=&\frac{\frac{h_{AA}^{a}}{f_{AA}^{a}}\frac{\left\langle N_{A}\right\rangle ^{2}}{N_{0}^{2}}}{1+\frac{h_{AA}^{a}}{f_{AA}^{a}}\frac{\left\langle N_{A}\right\rangle ^{2}}{N_{0}^{2}}+\frac{h_{BB}^{a}}{f_{BB}^{a}}\frac{\left\langle N_{B}\right\rangle^2 }{N_{0}^2}}\nonumber\\
  a_{\emph{bare}}&=&\frac{1}{1+\frac{h_{AA}^{a}}{f_{AA}^{a}}\frac{\left\langle N_{A}\right\rangle ^{2}}{N_{0}^{2}}+\frac{h_{BB}^{a}}{f_{BB}^{a}}\frac{\left\langle N_{B}\right\rangle^2 }{N_{0}^2}}\nonumber\\
  a_{\emph{off}}&=&\frac{\frac{h_{BB}^{a}}{f_{BB}^{a}}\frac{\left\langle N_{B}\right\rangle^2 }{N_{0}^2}}{1+\frac{h_{AA}^{a}}{f_{AA}^{a}}\frac{\left\langle N_{A}\right\rangle ^{2}}{N_{0}^{2}}+\frac{h_{BB}^{a}}{f_{BB}^{a}}\frac{\left\langle N_{B}\right\rangle^2 }{N_{0}^2}}\nonumber\\
  b_{\emph{on}}&=&\frac{\frac{h_{A}^{b}}{f_{A}^{b}}\frac{\left\langle N_{A}\right\rangle }{N_{0}}}{1+\frac{h_{A}^{b}}{f_{A}^{b}}\frac{\left\langle N_{A}\right\rangle }{N_{0}}}\nonumber\\
  b_{\emph{bare}}&=&\frac{1}{1+\frac{h_{A}^{b}}{f_{A}^{b}}\frac{\left\langle N_{A}\right\rangle }{N_{0}}},\label{eq19}
\end{eqnarray}
using $\sum_i a_i=1=\sum_j b_j$. If the genes change their state fast enough as compared to other time scales in the system, they will reach the stationary values of Eq.~(\ref{eq19}) before the other processes are completed; therefore they may be used in equations like (\ref{eq13}--\ref{eq17}). Now we are prepared to discuss the different limiting cases of fast, slow and ultra-slow genes.

\subsection{Fast genes}\label{secIII1}
In the limit of fast genes  we choose all binding and unbinding rates a hundred times larger than the decay rate of the fast protein $A$, that is $\delta^A$. In this limit the proteins see only average values of gene expression patterns, averaged over the five gene states. Therefore, in this limit we sum Eqs.~(\ref{eq13}), (\ref{eq14}), (\ref{eq15}) to predict the time evolution of $N_A$, and Eqs.~(\ref{eq16}) and (\ref{eq17}) for the time evolution of $N_B$ to obtain
\begin{eqnarray}
  \frac{d\Phi_{A}}{dt}&=&\frac{g_{\emph{bare}}^{a}+g_{\emph{on}}^{a}x_{AA}^{a} \Phi_{A}^{2}+g_{\emph{off}}^{a}x_{BB}^{a} \Phi_{B}^2}{1+x_{AA}^{a}\Phi_{A}^{2}+x_{BB}^{a}\Phi_{B}^2}-\delta^{A}\Phi_{A}\label{eqdetfastgenes1}\\
  \frac{d \Phi_{B}}{dt}&=&\frac{g_{\emph{bare}}^{b}+g_{\emph{on}}^{b}x_{A}^{b}\Phi_{A} }{1+x_{A}^{b} \Phi_{A}}-\delta^{B} \Phi_{B},\label{eqdetfastgenes}
\end{eqnarray}
where we have defined $\langle N_S\rangle/N_0\equiv\Phi_S$, $S=A,B$, and $x_{n}^{m}=\frac{h_{n}^{m}}{f_{n}^{m}}$ with $m$ referring to the respective gene and $n$ indicating the monomer or dimer binding of the transcription factors according to the chosen Hill coefficient in the biochemical reactions.

\subsubsection{Comparison with the deterministic description of a bistable frustrated unit}\label{secIII11}
Let us first briefly compare the equations~(\ref{eqdetfastgenes1},\ref{eqdetfastgenes}) with the deterministic equations formerly used to describe the bistable frustrated unit in \cite{bartek,pablohmo}
\begin{eqnarray}
  \frac{d\Phi_A}{dt} &=& \frac{\alpha}{1+(\Phi_B/K)}\;\cdot\;(\frac{b+\Phi_A^2}{1+\Phi_A^2})\;-\;\Phi_A\label{eqfirst}\\
  \frac{d\Phi_B}{dt} &=& \gamma (\Phi_A- \Phi_B),\label{eqsan}
\end{eqnarray}
where $\gamma$ is the ratio of the half-life of $\Phi_A$ to that of $\Phi_B$, that is $\delta^B/\delta^A$. In these units, the parameter $K$ sets the strength of repression of $\Phi_A$ by $\Phi_B$. The parameter $b$ determines the basal expression level of $A$, $b<1$.
The parameter $\alpha$ is the maximal rate of production of $A$ for full activation ($\Phi_A^2\gg b$) and no repression ($\Phi_B\approx 0$). If we divide Eqs.~(\ref{eqdetfastgenes1}~\ref{eqdetfastgenes}) by $\delta^A$ and define $\tau=t\delta^A$, $\gamma^a_{\emph{on}}=\frac{g^a_{\emph{on}}}{\delta^A}$, similarly for the other $g$-parameters, we have
\begin{eqnarray}
  \frac{d\Phi_{A}}{dt}&=&\frac{\gamma_{\emph{bare}}^{a}+\gamma_{\emph{on}}^{a}x_{AA}^{a} \Phi_{A}^{2}+\gamma_{\emph{off}}^{a}x_{BB}^{a} \Phi_{B}^2}{1+x_{AA}^{a}\Phi_{A}^{2}+x_{BB}^{a}\Phi_{B}^2}-\Phi_{A}\label{eq1a}\\
  \frac{d \Phi_{B}}{dt}&=&\frac{\delta^B}{\delta^A}\left(\frac{\gamma_{\emph{bare}}^{b}+\gamma_{\emph{on}}^{b}x_{A}^{b}\Phi_{A} }{1+x_{A}^{b} \Phi_{A}}-\Phi_{B}\right).\label{eqdetfastgenesrescaled}
\end{eqnarray}
In the previous model we used $\alpha$ as a bifurcation parameter which multiplies $\Phi_A^2$ in the gain term of Eq.~(\ref{eqfirst}), a similar role in Eq.~(\ref{eq1a}) is played by $g_{\emph{on}}^a$ which we therefore use here as the bifurcation parameter. Differently from our former parametrization, apart from the common prefactor $\delta^B/\delta^A$, that sets the slow time scale of $\Phi_B$, the gain term in the second equation~(\ref{eqdetfastgenesrescaled}) implicitly depends on $1/\delta^{B}$, compared to the loss term. Therefore to align the scale of production with the slow decay, we have to adjust the production by choosing $g^b_{\emph{on},\emph{bare}}$ each by two orders of magnitude smaller than the corresponding production rates of $g^a_{\emph{on},\emph{bare}}$ which explains our choice in table~\ref{table1}. So the slow dynamics of protein $B$ is realized by both slow decay and slow production rate on the genetic level.

\noindent
Moreover, it should be noticed that we have changed the power of the Hill coefficient in the binding term of the repressor concentration $\Phi_B$ from one in Eq.~(\ref{eqfirst}) to two in Eq.~(\ref{eq1a}) corresponding to the choice of $h^a_{BB}(i)\frac{N_B^2}{N_0^2}$ in the master equation (\ref{eqmaster}).
This appears as a minor difference in the equations. The effect, however, is a considerable broadening of the intermediate limit cycle regime in case of a Hill coefficient of $2$. Since we are interested in the fate of the regular oscillations in case of slow and ultra-slow genes, it is important not to need a finetuning for seeing oscillations for fast genes.

\noindent Furthermore, we would like to compare our equations (\ref{eq1a},\ref{eqdetfastgenesrescaled})
with the deterministic equations as they were derived for ``design I" in \cite{guantes}. In common with those equations of \cite{guantes} is the limit of fast genes and the realization of the repression operating on the transcriptional level. The main differences between both sets of equations are firstly the power one of $\Phi_A$ in Eq.~(\ref{eqdetfastgenesrescaled}), which can be traced back to the monomer (rather than dimer) binding of the transcription factor $A$ to the promoter region of gene $b$ in Eq.~\ref{eqmaster}, and secondly, the relative weights between gain and loss terms. In particular the bifurcation parameter affects in our case only the first equation directly and the second equation indirectly via $\Phi_A$, while it affects both equations directly in \cite{guantes}. The combination of these apparently minor differences leads to different bifurcation patterns: a saddle-node bifurcation in \cite{guantes} and Hopf bifurcations in our case. When increasing our bifurcation parameter $g^a_{\emph{on}}$, we see two fixed-point regimes for low and high values of $g^a_{\emph{on}}$, separated by an intermediate limit-cycle regime due to two Hopf bifurcations. Therefore, the deterministic equations (\ref{eq1a},\ref{eqdetfastgenesrescaled}) reproduce the phase structure of the bistable frustrated unit. Naively one may expect that the actual bifurcation scenarios in the deterministic limit are irrelevant for the final stochastic systems. It is, however, known from the work of \cite{izhikevich}
in the context of neural networks and also emphasized in \cite{guantes} that the very bifurcation scenario may have a strong impact on the final biological function of the motif, as the very onset of oscillations and the embedding in phase space have an impact on amplitude, frequency, noise resistance and other stability properties. Therefore we present a detailed bifurcation analysis of Eqs.~(\ref{eq1a},\ref{eqdetfastgenesrescaled}) in the Appendix. As it is seen there, the analysis requires a further zoom into the two transition regions, that is, a high resolution and finetuning of the bifurcation parameter. It would be interesting to search for manifestations or remnants of these scenarios in a fully stochastic description, of which we studied so far the gross features only: ``noisy" fixed points and ``noisy" limit cycles.

\subsubsection{Gillespie simulations for fast genes}\label{secIII12}

\begin{figure}[h]
  \begin{centering}
    \includegraphics[width=6.5cm]{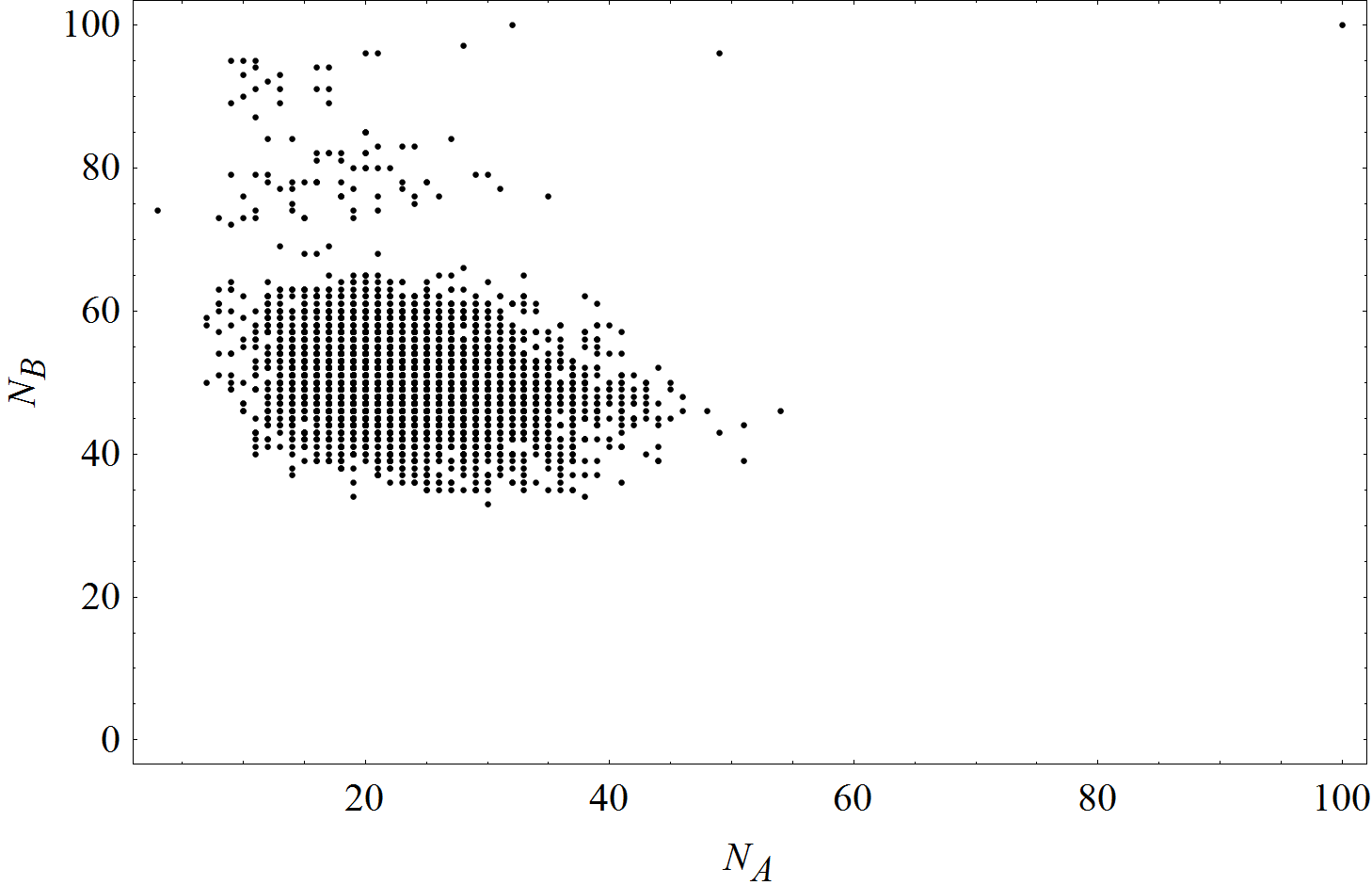}$\;$$\;$
    \includegraphics[width=6.5cm]{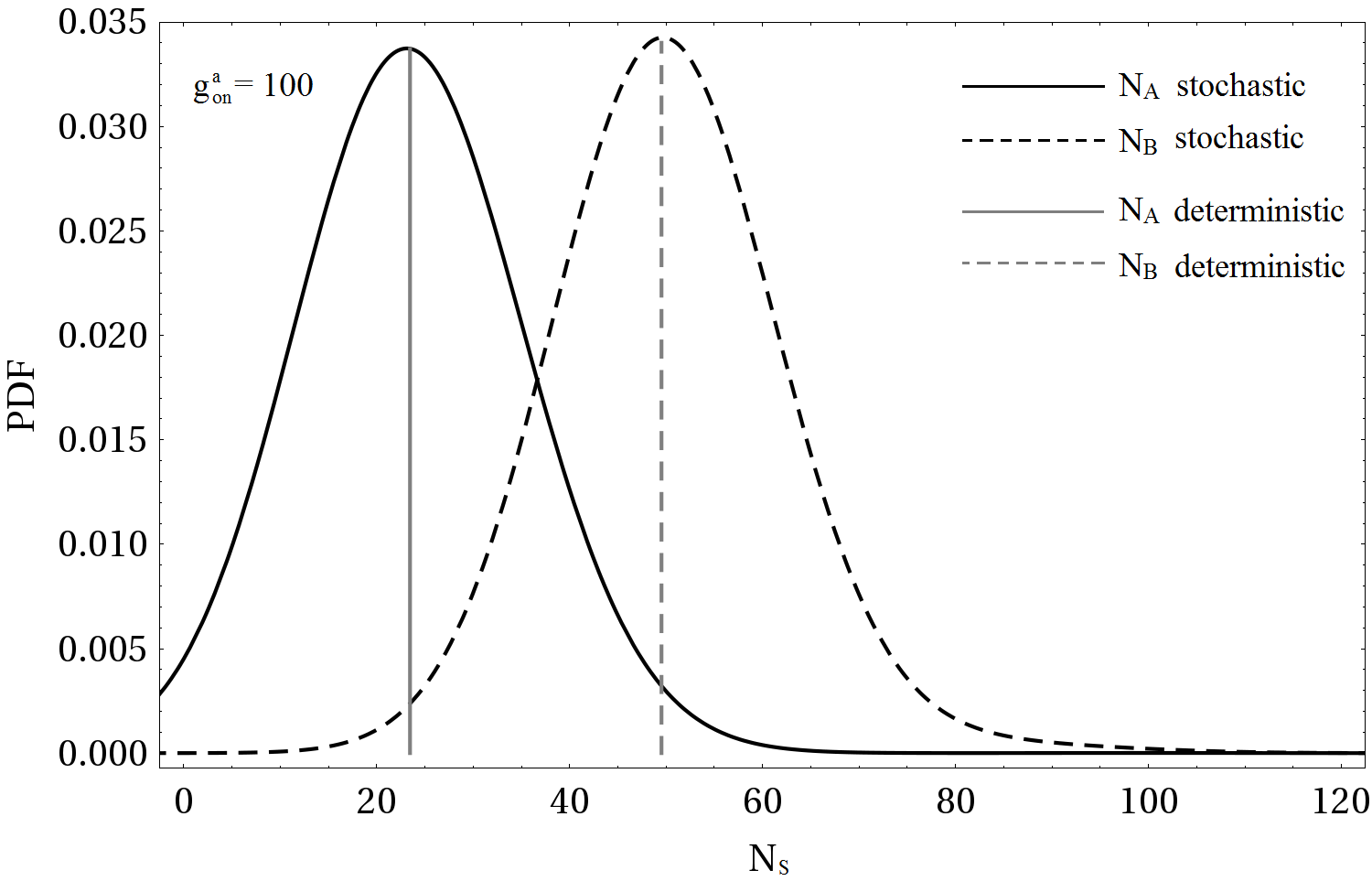}
    \par\end{centering}

  \begin{centering}
    \includegraphics[width=6.5cm]{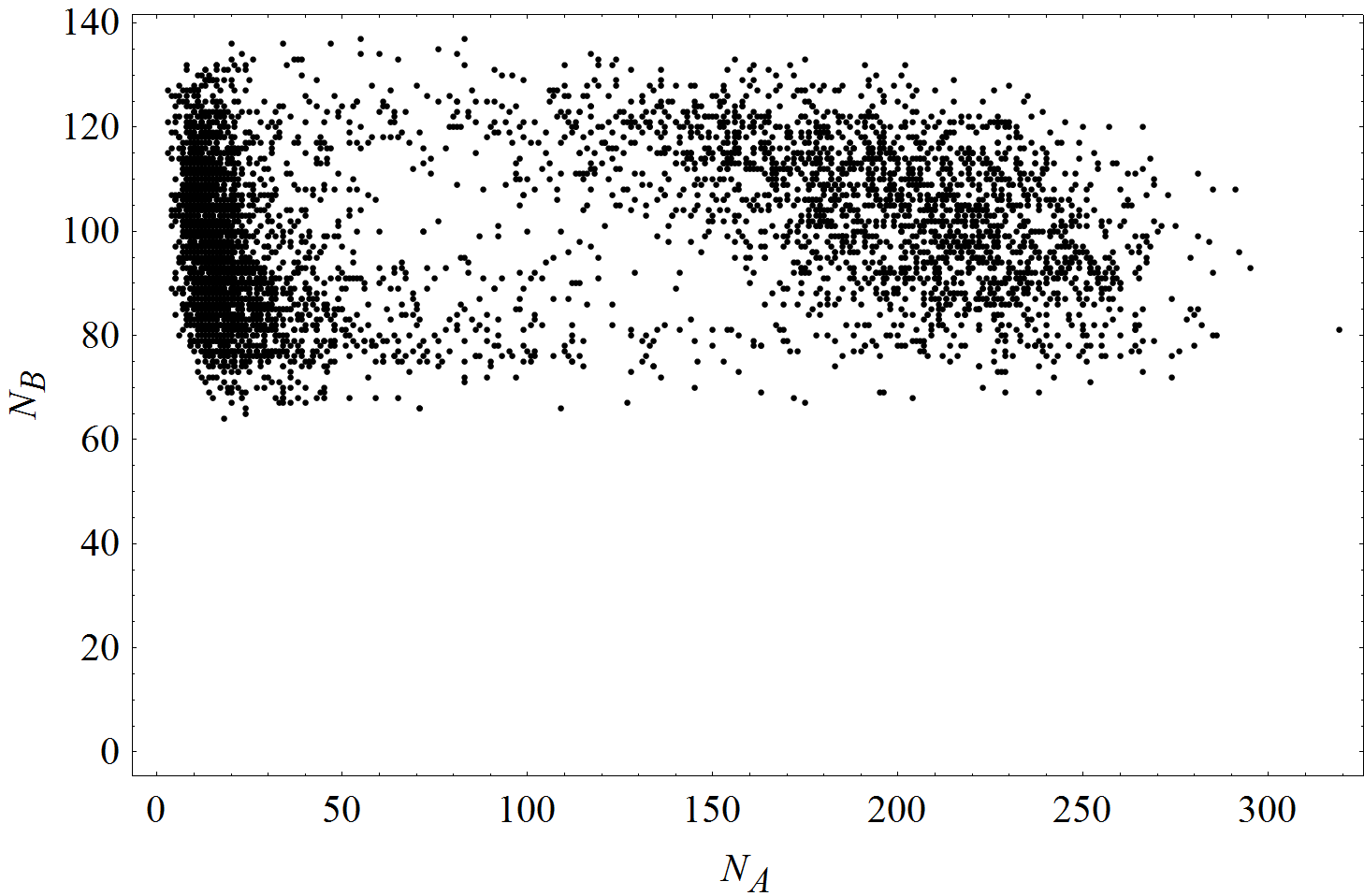}$\;$$\;$
    \includegraphics[width=6.5cm]{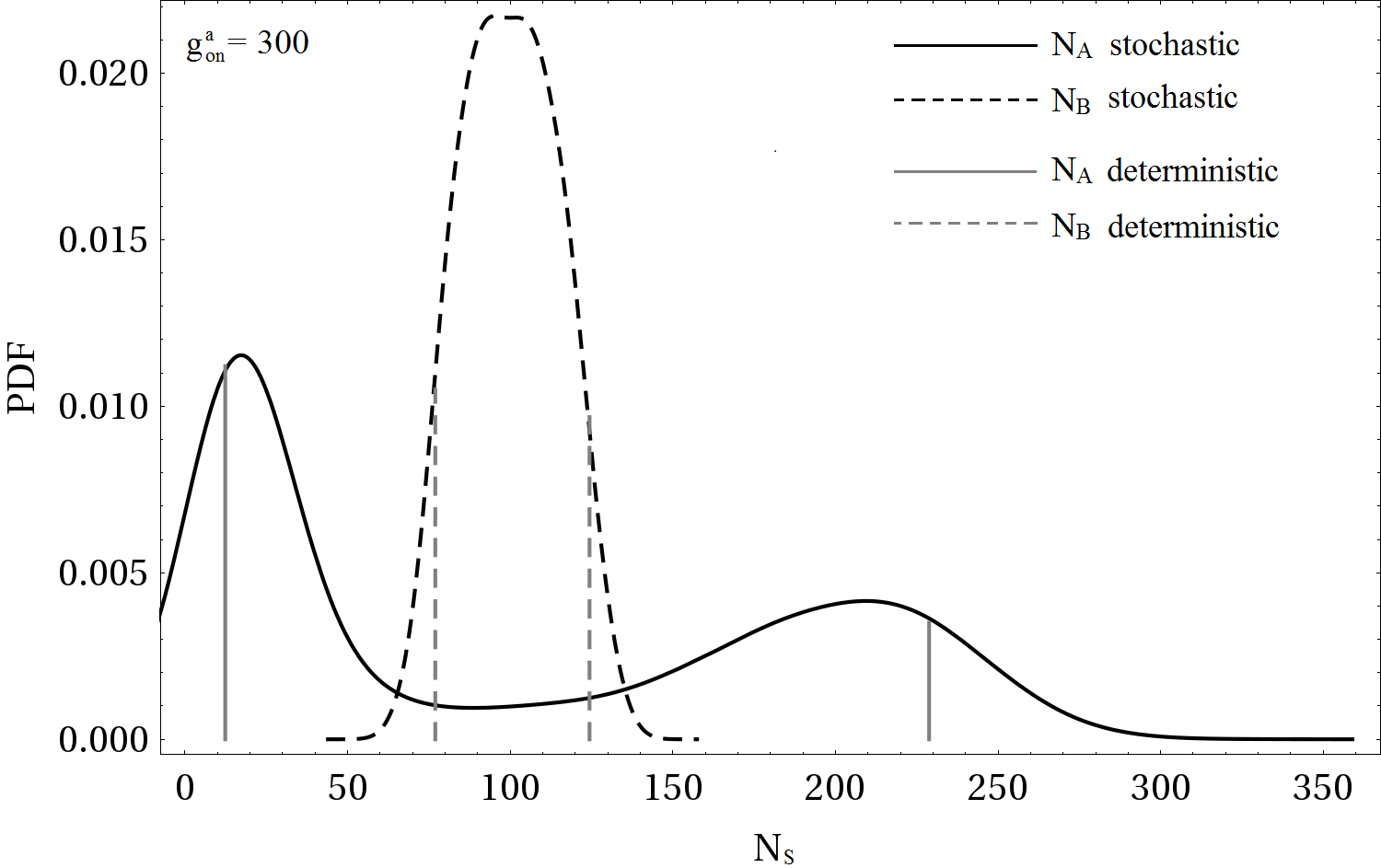}
    \par\end{centering}

  \begin{centering}
    \includegraphics[width=6.5cm]{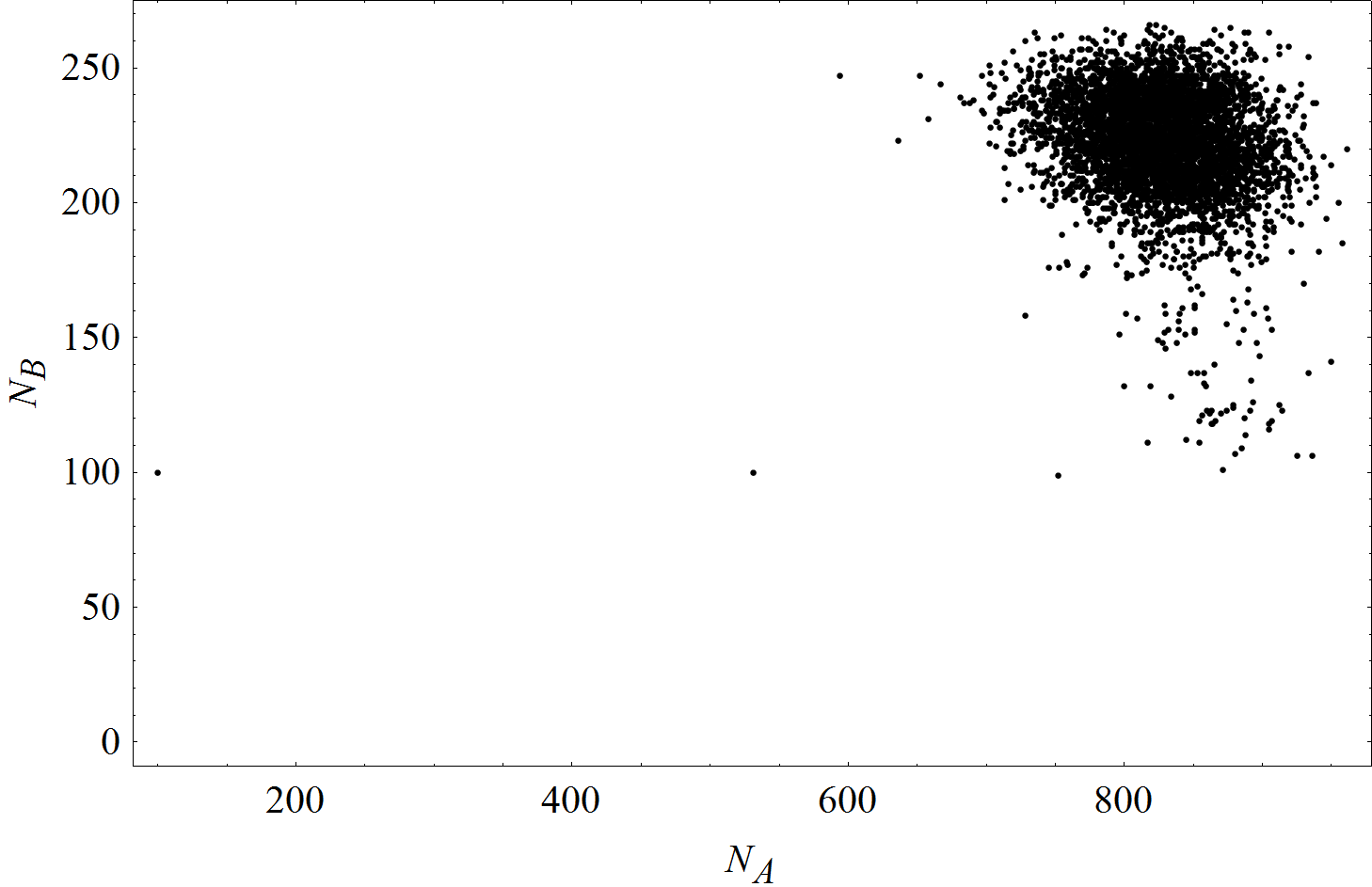}$\;$$\;$
    \includegraphics[width=6.5cm]{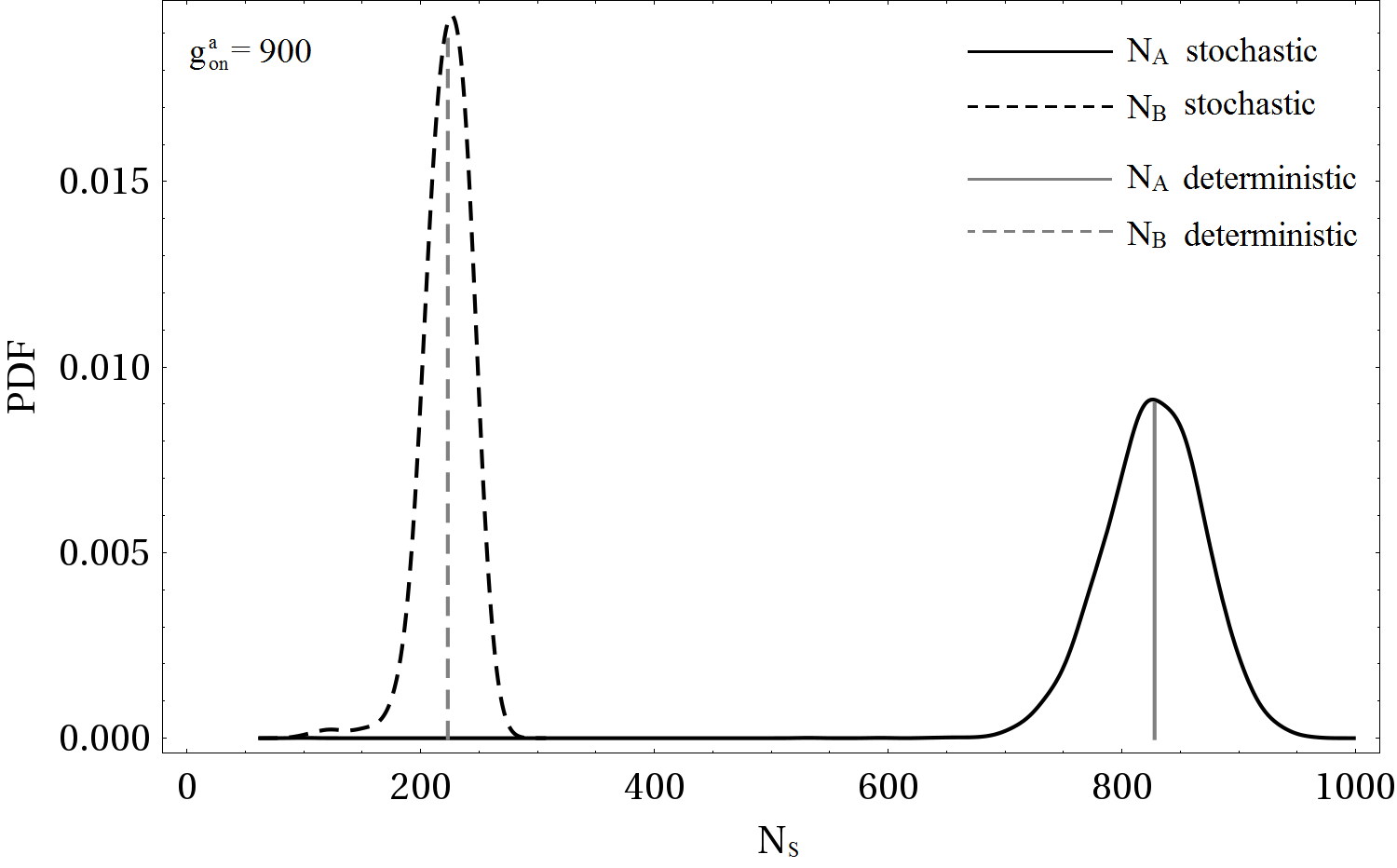}
    \caption{Gillespie simulations for fast genes, that is $h^a_{AA}=h^a_{BB}=0.01$, $h^b_A=1.0$, $f^a_{AA}=f^a_{BB}=f^b_A=100$. Phase portraits of the number of proteins $N_B$ versus $N_A$ within Gillespie time $T_G=5000$ (left column) and corresponding probability density functions (PDF) (right column) of $N_A$ (full line black) and $N_B$ (dashed line black), while the gray full and dashed vertical lines indicate the position of the fixed points in the first and third row. In the first and third row $g^a_{\emph{on}}=100$ and $g^a_{\emph{on}}=900$, respectively, in the left panels we see the stochastic pendant of the fixed points observed in the deterministic case. For clarity of the figure we do not plot every Gillespie step, but only $5000$ of them. The maxima of the PDFs agree well with the fixed points in the deterministic description. In the second row we see the stochastic version of limit cycles for $g^a_{\emph{on}}=300$. The vertical lines here mark the maximal and minimal extension of the limit cycles in $\Phi_A$ and $\Phi_B$ when integrated as solutions of the deterministic equations (\ref{eqdetfastgenes1}),(\ref{eqdetfastgenes}).  These plots confirm our former model (\ref{eqfirst}),(\ref{eqsan}) as a suitable coarse-grained description.}\label{fig11}
  \end{centering}
\end{figure}

We present results of Gillespie simulations of the reactions, listed in Eqs.~(\ref{eq4}) and (\ref{eq3}) for parameter values as in Table~\ref{table1} and the first row of Table~\ref{table2}. Figure~\ref{fig11}, left column, shows phase portraits of $N_B$ versus $N_A$ for three values of the bifurcation parameter, $g^a_{\emph{on}}=100$ (first row), $g^a_{\emph{on}}=300$ (second row) and $g^a_{\emph{on}}=900$ (third row), which are typical for the lower ($g^a_{\emph{on}}=100$) and higher ($g^a_{\emph{on}}=900$) ``fixed-point regime", and for  the ``limit-cycle regime" ($g^a_{\emph{on}}=300$). The phase portraits are made within a Gillespie time of $T_G=\sum_i\;dt_i=5000$, where $dt_i$ refers to the time interval, randomly chosen out of a Poisson distribution in the $i$-th Gillespie update. The regimes are termed after their deterministic pendants: In the deterministic limit ($N_0\rightarrow\infty$) the clouds of $N_A,N_B$ values in the first and third row would shrink to the lower and higher fixed points as predicted from Eqs.~(\ref{eqdetfastgenes1},\ref{eqdetfastgenes}), while the clouds in the second row would contract to a limit cycle. (The higher density of $(N_A,N_B)$-values for small and large values of $N_A$ is due to the fact that also the stochastic version of a limit cycle spends more time in regions, where $N_B$ drastically changes, since $B$ is the slow variable, while large changes in $N_A$ happen rapidly, since $A$ is the fast variable.) The probability density functions in the right column of the figures reflect the probability for finding concrete combinations of $(N_A,N_B)$-values in the phase portraits. They are displayed for a quantitative comparison of their maxima with the prediction of the location of the fixed points and the extension of the limit cycle from the deterministic equations. These locations are indicated via the vertical lines. The vertical lines hit the maxima quite well where they correspond to fixed points; the lines also match the typical extension of the cloud in case of the limit cycle.

\begin{figure}
  \begin{centering}
    \includegraphics[width=8cm]{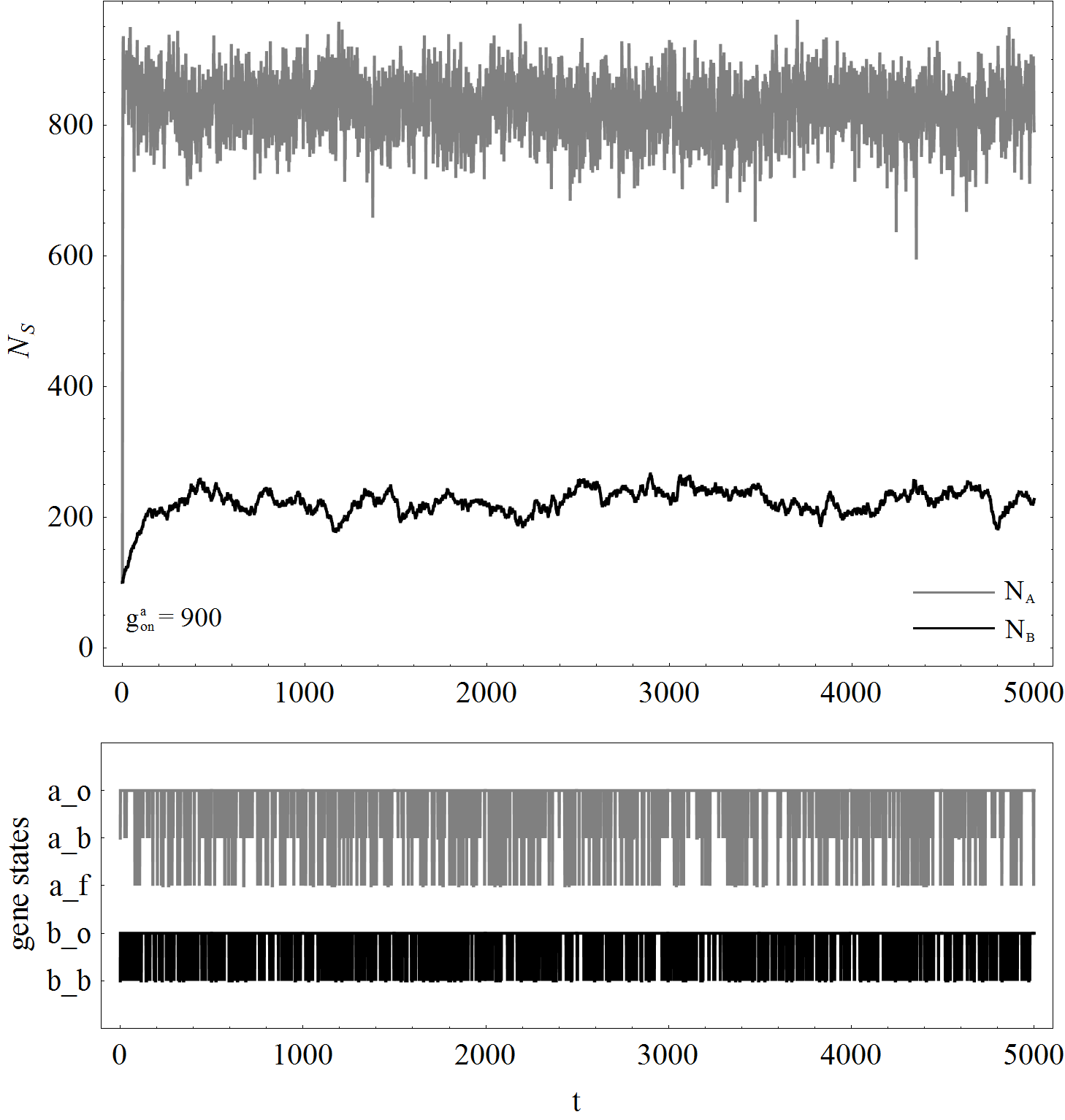}
    \par\end{centering}
  \caption{Time series of the number of protein species $S$, $S=A$~(gray), $S=B$~(black) (upper part of the figure) and the gene states for gene~$a$ (gray) and gene~$b$ (black) (lower part of the figure), recorded during the Gillespie steps. The bifurcation parameter is chosen as $g^a_{\emph{on}}=900$. States of gene $a$ switch between $\emph{on}$, [$a_{o}$ (gray)], $\emph{bare}$, [$a_b$ (white)], $\emph{off}$, [$a_f$ (gray)], states of gene $b$ switch between $\emph{on}$, [$b_o$ (black)] and $\emph{bare}$, [$b_b$ (white)]. The other parameters are chosen as $g_{\emph{bare}}^{a}= 25$, $g_{\emph{off}}^{a}= 0$, $g_{\emph{on}}^{b} = 2.5$, $g_{\emph{bare}}^{b} = 0.025$, $h_{AA}^{a}$ = $h_{BB}^{a} = 0.01$, $h_{A}^{b} = 1$, $f_{AA}^{a}=$ $f_{BB}^{a}=$ $f_{A}^{b}$ $= 100$, $\delta^{A}= 1$, $\delta^{B} = 0.01$.}\label{fig10}
\end{figure}

\begin{figure}
  \begin{centering}
    \includegraphics[width=6.5cm]{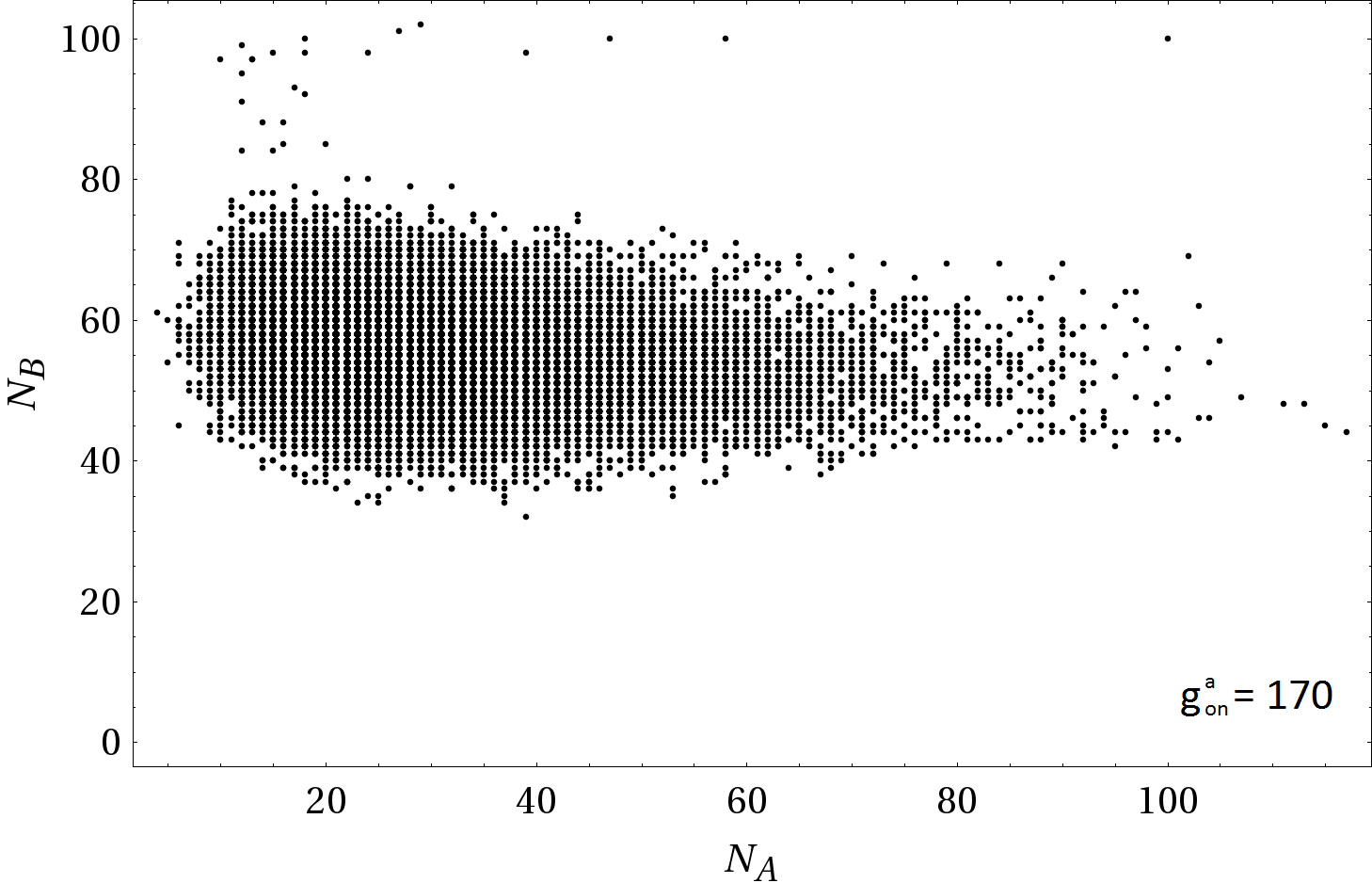}$\;$$\;$
    \includegraphics[width=6.5cm]{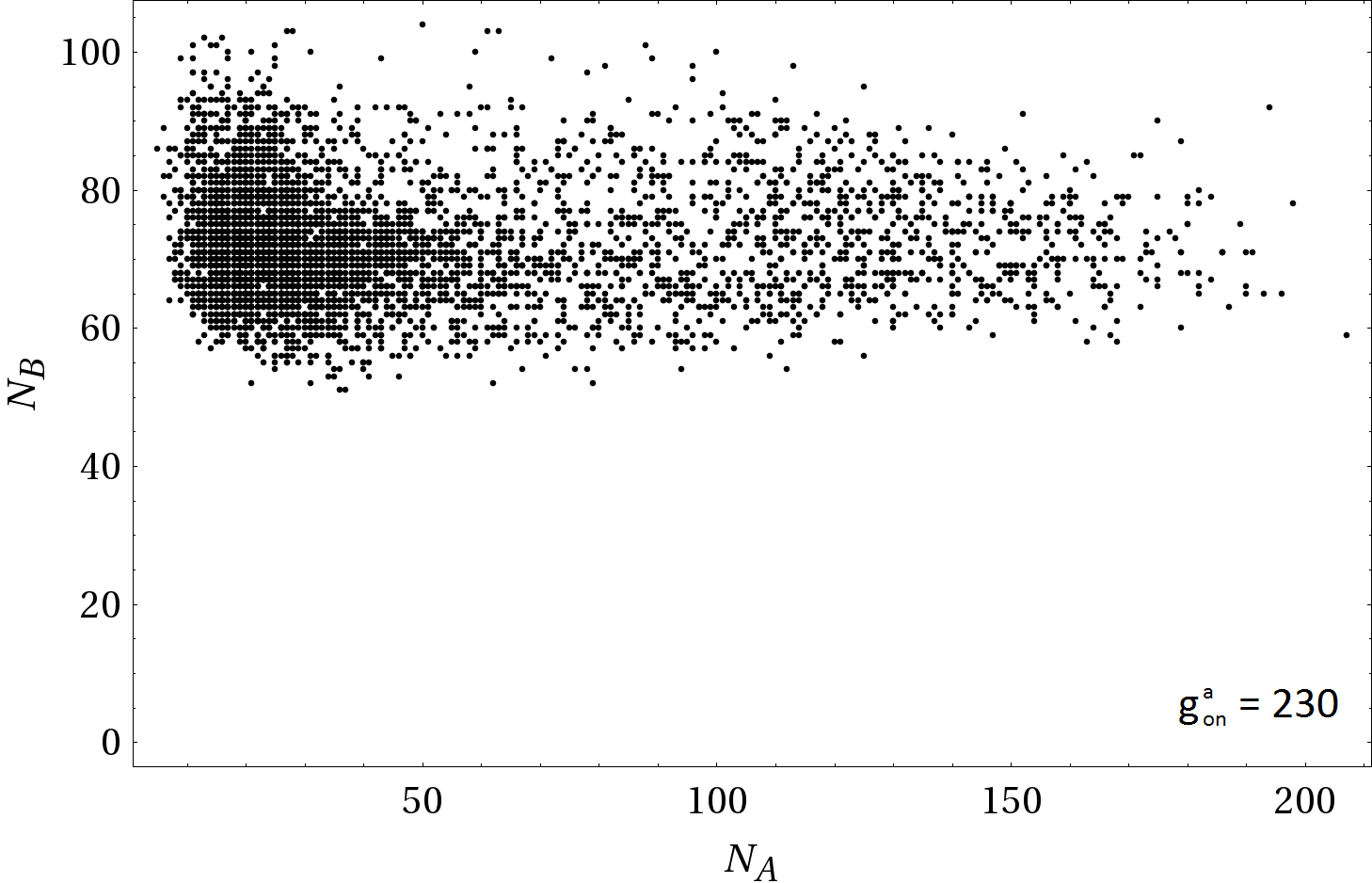}

    \begin{centering}
      \includegraphics[width=6.5cm]{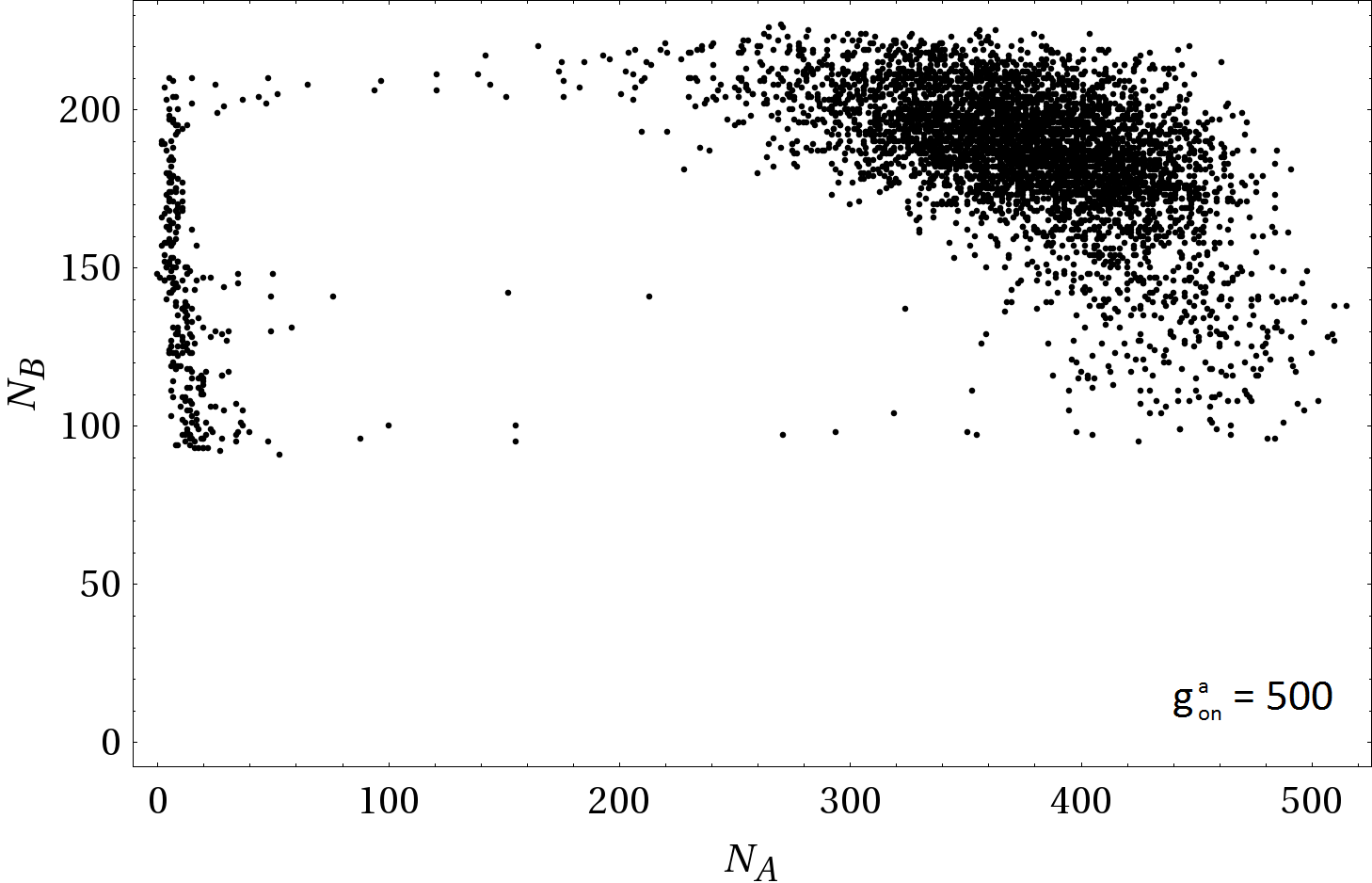}$\;$$\;$
      \includegraphics[width=6.5cm]{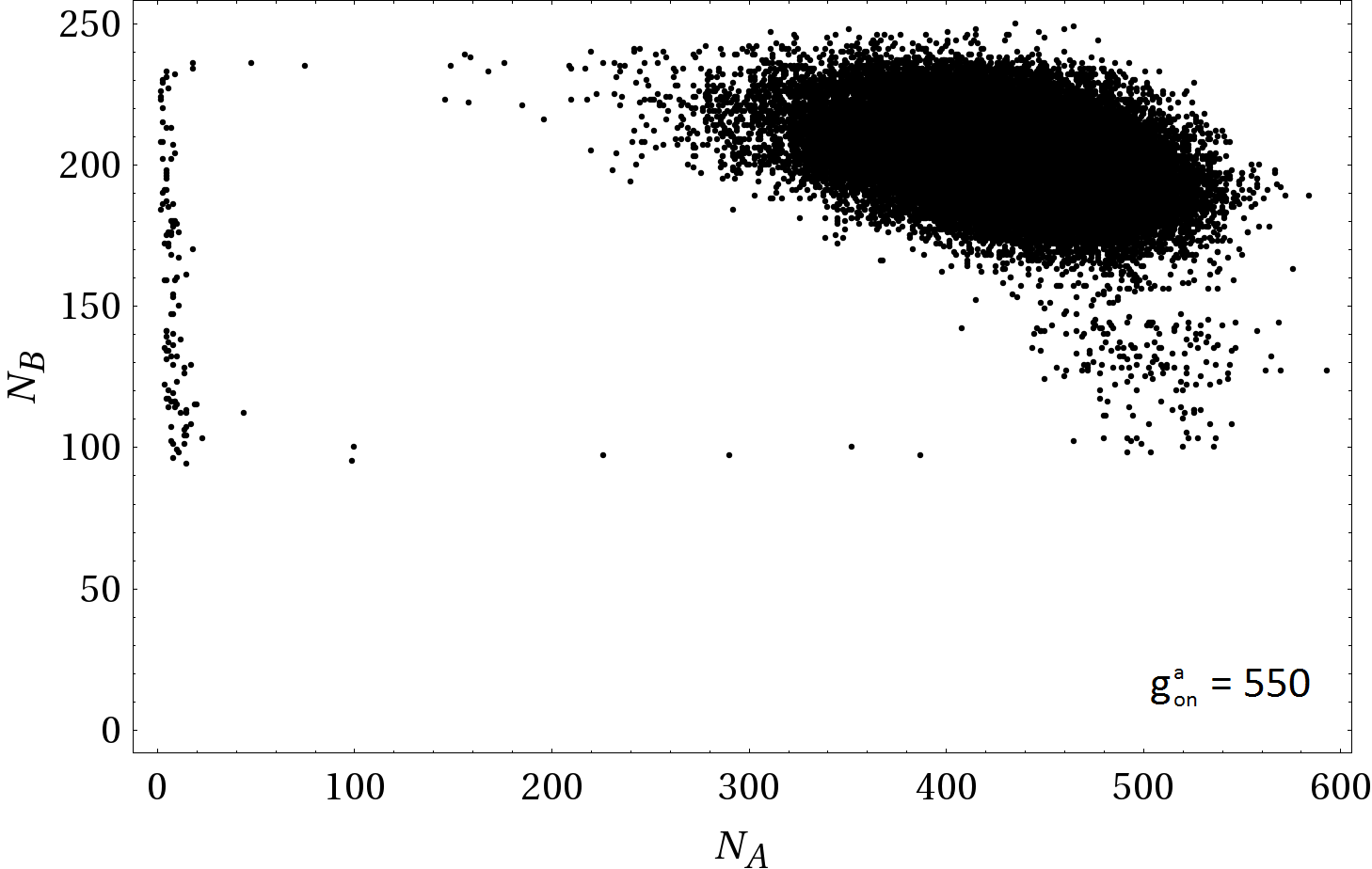}
      \par\end{centering}
    \caption{Phase portraits within Gillespie time $T_G=10000$ for fast genes for four values of the bifurcation parameter. $g^a_{\emph{on}}=170$ (upper left), $g^a_{\emph{on}}=230$ (upper right), $g^a_{\emph{on}}=500$ (lower left), $g^a_{\emph{on}}=550$ (lower right). In the lower ($g^a_{\emph{on}}=170,230$) and higher ($g^a_{\emph{on}}=500,550$) fixed point regimes we see clear indications of quasi-cycles in the stochastic descriptions.}\label{fig12}
  \end{centering}
\end{figure}

\noindent Figure~\ref{fig10} (upper part) shows the time series of the number of proteins~$A$, $N_A$ (gray) and of proteins~$B$, $N_B$ (black), which are fluctuating about constant values $(N_A\sim 800,N_B\sim 200)$. The fluctuations between the different gene states, in particular between the on- and bare-states, is so fast that it appears as a gray and black band (Fig.~\ref{fig10} lower part), so that the protein values, shown in the upper part of the figure, only fluctuate about the fixed point values.
In the Gillespie simulations of our former realization of the genetic circuit \cite{bartek} we identified quasi-cycles deeply in the fixed-point regimes. Such cycles, caused by large demographic fluctuations, are also found in the present realization of the genetic circuit if we wait sufficiently long for such fluctuations to happen. Figure~\ref{fig12} shows a few such excursions below (upper left figure for $g^a_{\emph{on}}=170$) and above (lower figures for $g^a_{\emph{on}}=500$ and $550$) the limit-cycle regime, while the data in the upper right figure for $g^a_{\emph{on}}=230$ are compatible with more frequent and smaller excursions into phase space due to the vicinity of the transition region to the region of regular limit cycles.

\subsection{Slow genes}\label{secIII2}

\begin{figure}
  \begin{centering}
    \includegraphics[width=9.cm]{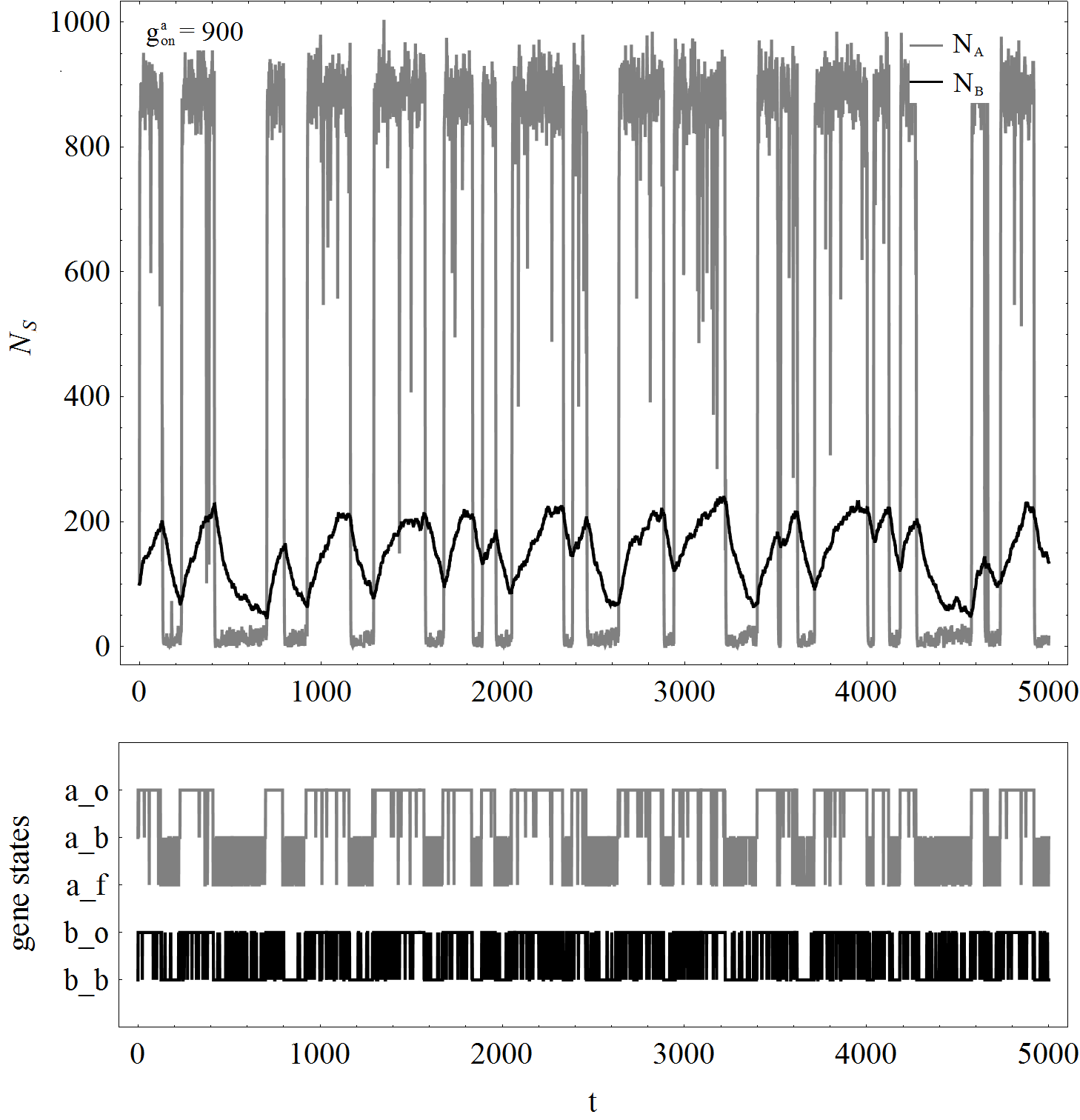}
    \par\end{centering}
  \caption{Same as Fig.~\ref{fig10}, but for slow genes, that is $h^a_{AA}=h^a_{BB}=0.0001$, $h^b_A=0.01$, $f^a_{AA}=f^a_{BB}=f^b_A=1.0$. Here the switching between $\emph{on}$ and $\emph{bare}$ states is much faster than the indirect switching between $\emph{on}$ and $\emph{off}$ states. The other parameters are as in Fig.~\ref{fig10}. }\label{fig13}
\end{figure}

In the limiting case of slow genes, the binding and unbinding rates are chosen to be of the order of the decay rate of the fast protein, so that they are still fast as compared to the decay rate of the slow protein $B$. Therefore we sum the moments of $N_B$ from Eqs.~(\ref{eq16}),(\ref{eq17}) also over the states of gene $b$. Moreover we have three equations (\ref{eq13}),(\ref{eq14}),(\ref{eq15}) that include already a summation over the states of gene $b$, but in order to further sum over states of gene $a$, we consider Fig.~\ref{fig13},
%%% FIGURE fig13
which shows the switching events of gene $a$ between the three states $\emph{on}, \emph{bare}, \emph{off}$ and of gene $b$ between the two states $\emph{on}$ and $\emph{bare}$. The switching events between the on- and the bare-state, determined by $h^a_{AA}$, $f^a_{AA}$, as well as the switching between the $\emph{bare}$- and the $\emph{off}$-state, determined by $h^a_{BB}$, $f^a_{BB}$, are still frequent as compared to an indirect switching between an $\emph{on}$- and an $\emph{off}$-state of gene $a$, since it is due to different binding events. Therefore the proteins $A$ and $B$ effectively see two gene levels, an average over the on- and bare-state, or over the bare- and off-state. These two different levels are seen in the time evolution of $N_A$: $N_A$ switches between the upper gray band in Fig.~\ref{fig13}, that is between $N_A=800$ and $N_A=1000$, and the lower gray band close to zero. The adaptation of $N_B$ to the abrupt changes in $N_A$ is more smooth and happens with delay. (As we shall later see, the fast proteins $A$ are able to adapt to these two states, while the slow ones $B$ are not.) Accordingly we average Eq.~(\ref{eq13}) with Eq.~(\ref{eq15}) and Eq.~(\ref{eq14}) with Eq.~(\ref{eq15}) to obtain two sets of differential equations
\begin{eqnarray}
  \frac{d\langle N_{A}\rangle}{dt}&=&N_{0}  \frac{g_{\emph{bare}}^{a}+g_{\emph{on}}^{a}x_{AA}^{a}\frac{\langle N_{A}\rangle ^{2}}{N_{0}^{2}}}{1+x_{AA}^{a}\frac{\langle N_{A}\rangle ^{2}}{N_{0}^{2}}}-\delta^{A}\langle N_{A}\rangle,\label{eq24a} \\
  \frac{d\langle N_{B}\rangle}{dt}&=&N_{0}  \frac{g_{\emph{bare}}^{b}+g_{\emph{on}}^{b}x_{A}^{b}\frac{\langle N_{A}\rangle }{N_{0}}}{1+x_{A}^{b}\frac{\langle N_{A}\rangle }{N_{0}}}-\delta^{B}\langle N_{B}\rangle\label{eq24b}
\end{eqnarray}
and
\begin{eqnarray}
  \frac{d\langle N_{A}\rangle}{dt}&=&N_{0}  \frac{g_{\emph{bare}}^{a}+g_{\emph{off}}^{a}x_{BB}^{a}\frac{\langle N_{B}^2\rangle }{N_{0}^2}}{1+x_{BB}^{a}\frac{\langle N_{B}^2\rangle }{N_{0}^2}}-\delta^{A}\langle N_{A}\rangle \label{eq25a}\\
  \frac{d\langle N_{B}\rangle}{dt}&=&N_{0}  \frac{g_{\emph{bare}}^{b}+g_{\emph{on}}^{b}x_{A}^{b}\frac{\langle N_{A}\rangle }{N_{0}}}{1+x_{A}^{b}\frac{\langle N_{A}\rangle }{N_{0}}}-\delta^{B}\langle N_{B}\rangle\label{eq25b},
\end{eqnarray}\enlargethispage{\baselineskip}
which describe the two alternatives of the time evolution of the system, either following the first or the second set of equations. While the time dependence of $\langle N_B\rangle$ is the same in both equations (\ref{eq24b}) and (\ref{eq25b}), the time dependence of $\langle N_A\rangle$ depends on whether it is due to the activating binding of a dimer of $A$ or the repressing binding of $B$. This apparently minor detail leads to a very different bifurcation pattern, as it is revealed by a linear stability analysis. For Eqs.~(\ref{eq25a}),(\ref{eq25b}) this analysis shows that the determinant of the Jacobian, evaluated at the single fixed point, is positive for all combinations of positive values of parameters and $g^a_{\emph{bare}}>g^a_{\emph{off}}$ and $g^b_{\emph{on}}>g^b _{\emph{bare}}$, while the trace of the Jacobian, evaluated at the fixed point, is negative, so the fixed point is always stable. In contrast, for the first set of equations, Eqs.~(\ref{eq24a}),(\ref{eq24b}), we have for our usual choice of parameters (used in the Gillespie simulations) a single fixed point, but there is a combination of parameters for a very low value of $g^a_{\emph{bare}}$, for which we obtain three positive real fixed points; for example choosing $g^a_{\emph{on}}=300$, $g^a_{\emph{bare}}=5$ and the other parameters as before, we obtain two stable nodes with positive determinant, negative trace and negative eigenvalues at $(N_A^\ast=6.09, N_B^\ast=16.71)$, and at $(N_A^\ast=262.65, N_B^\ast=181.75)$, and a saddle with negative determinant, positive trace, and $\lambda_1<0$, $\lambda_2>0$ at $(N_A^\ast=31.26, N_B^\ast=61.44)$. Depending on the initial conditions, the system will evolve in one of the stable nodes. Gillespie simulations for this parameter set are not displayed in Fig.~\ref{fig14} below. The basin of attraction of the lower fixed point is so small that  the stochastic system practically never evolves to this fixed point due to the demographic fluctuations.

\subsubsection{Gillespie simulations for slow genes}\label{secIII21}

\begin{figure}
  \begin{centering}
    \includegraphics[width=6.5cm]{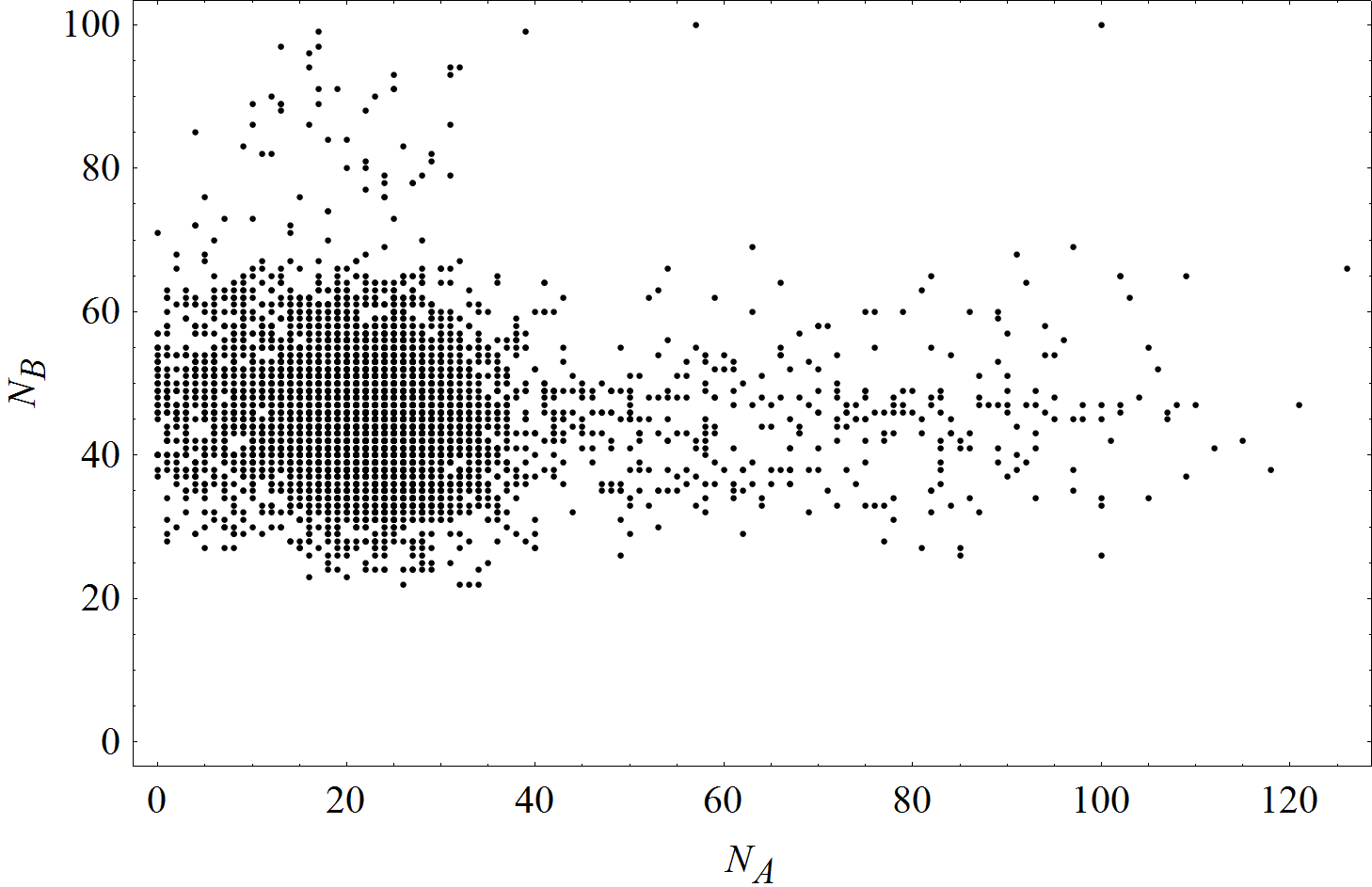}$\;$$\;$
    \includegraphics[width=6.5cm]{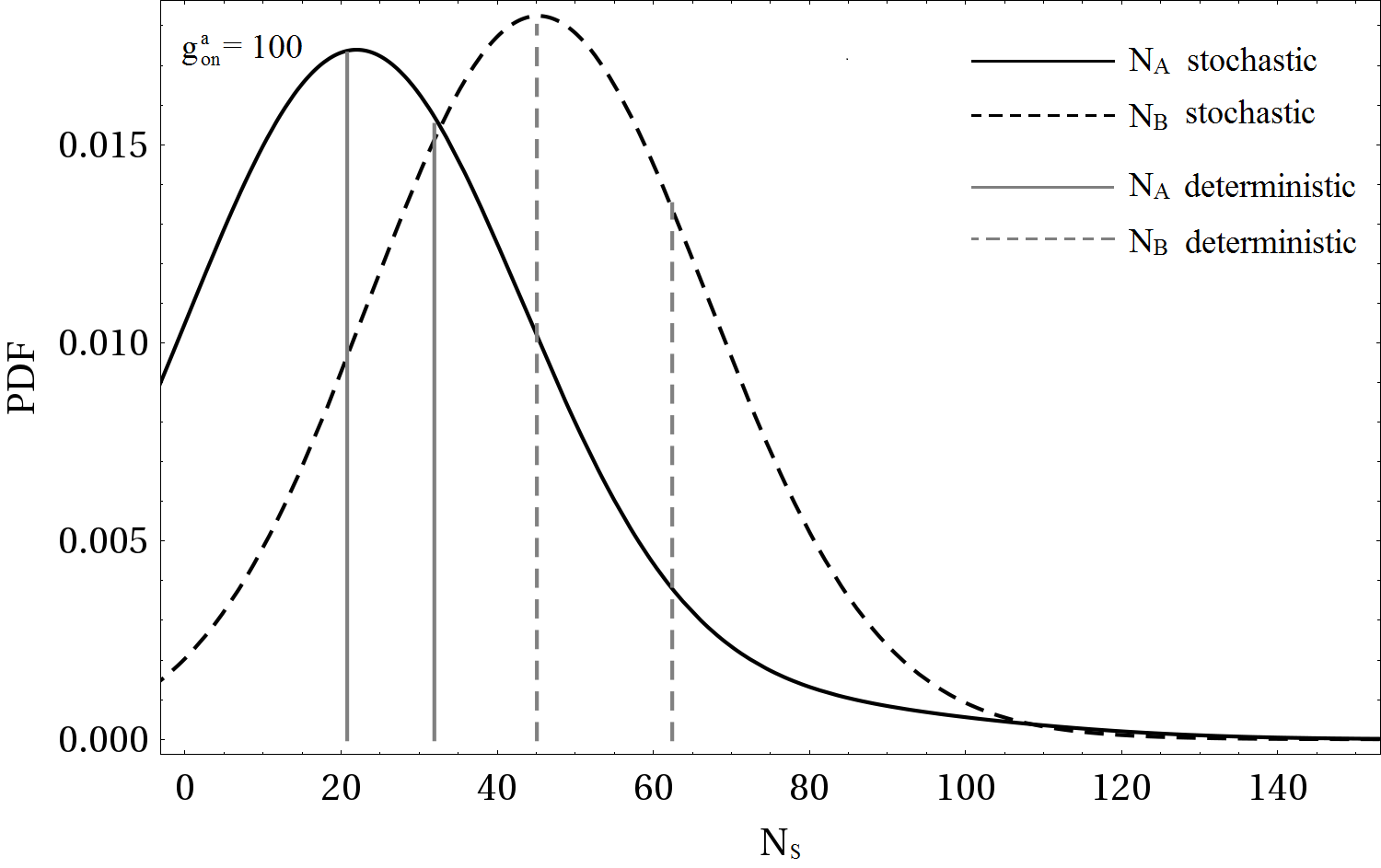}
    \par\end{centering}

  \begin{centering}
    \includegraphics[width=6.5cm]{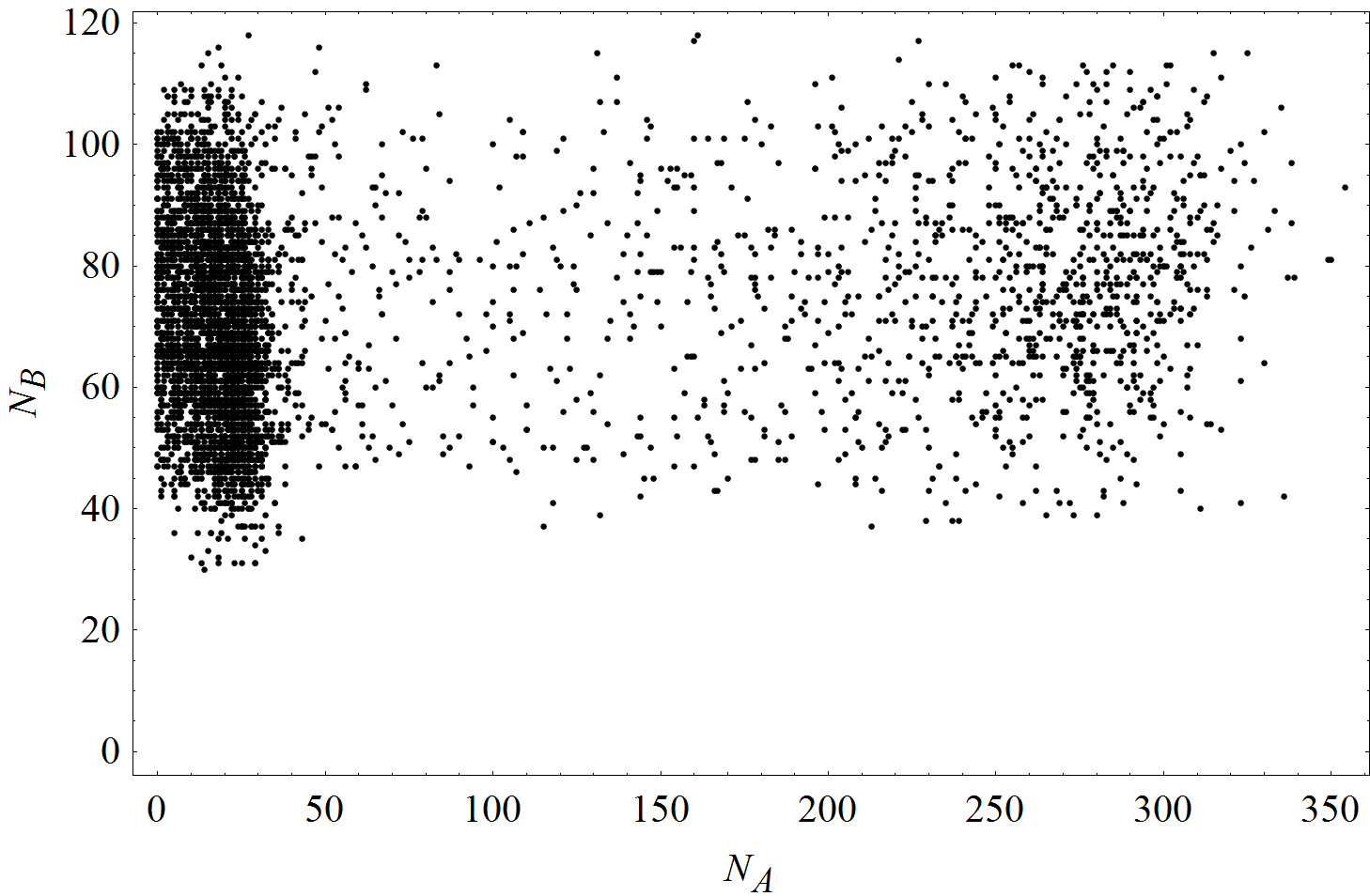}$\;$$\;$
    \includegraphics[width=6.5cm]{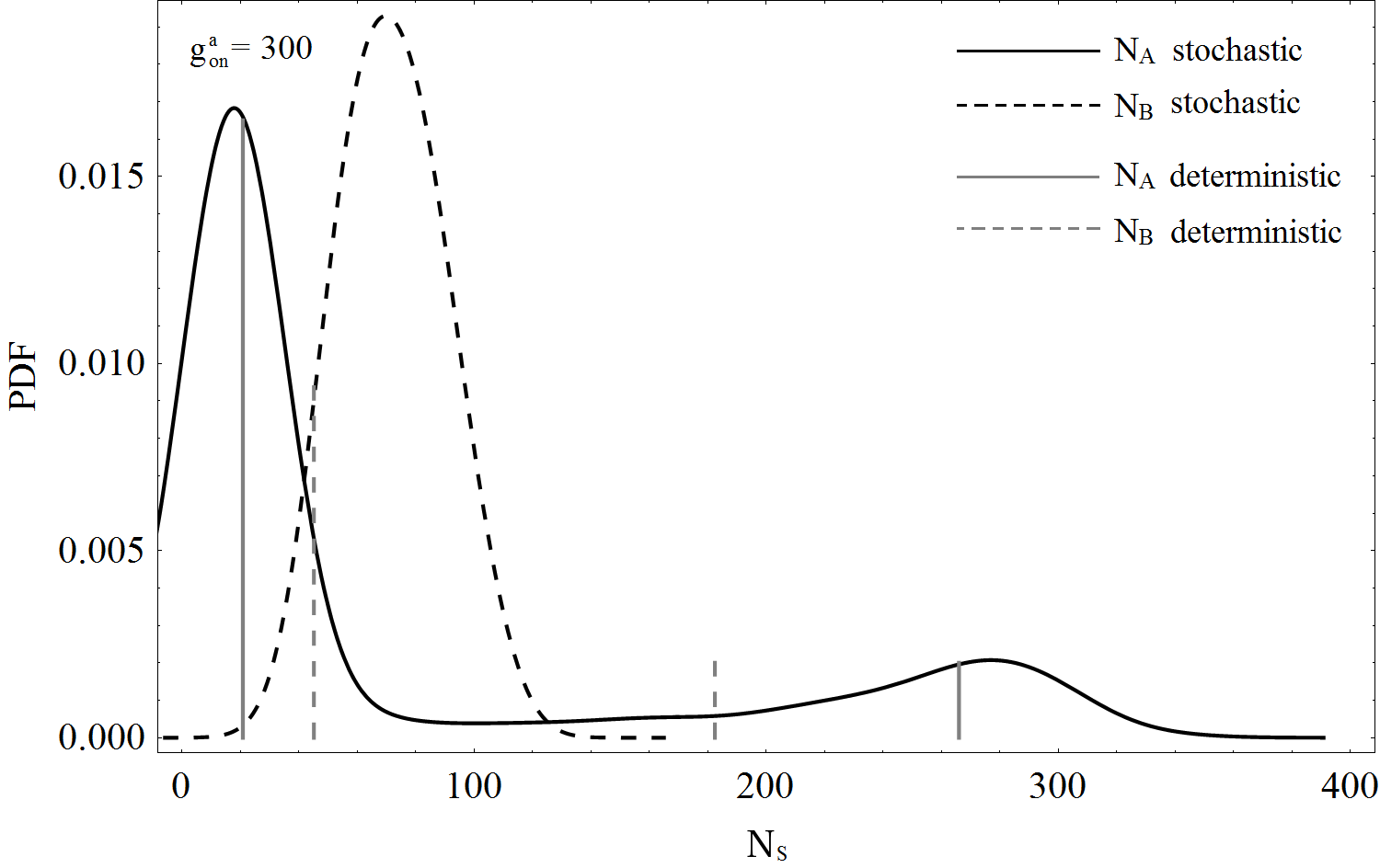}
    \par\end{centering}

  \begin{centering}
    \includegraphics[width=6.5cm]{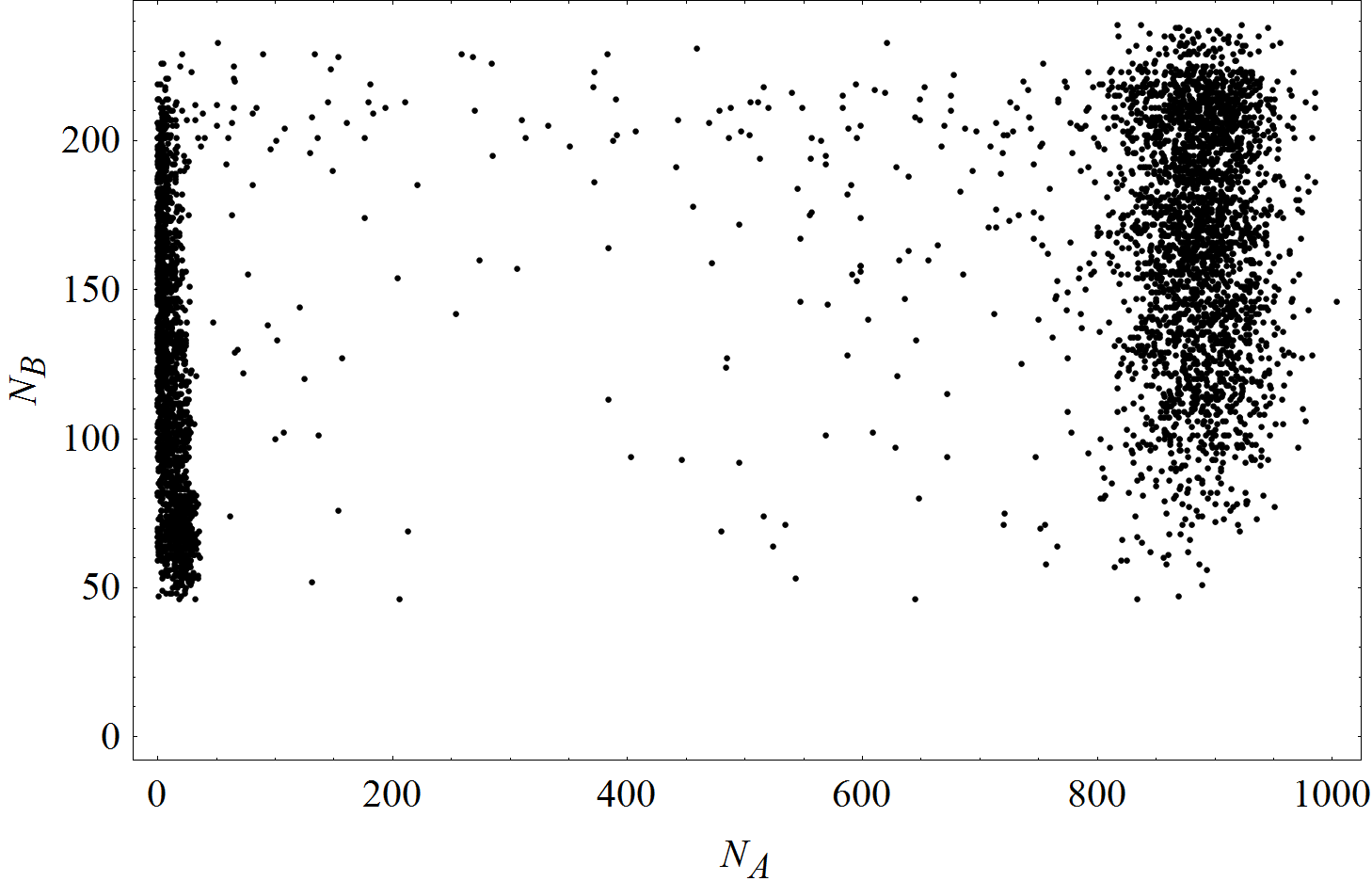}$\;$$\;$
    \includegraphics[width=6.5cm]{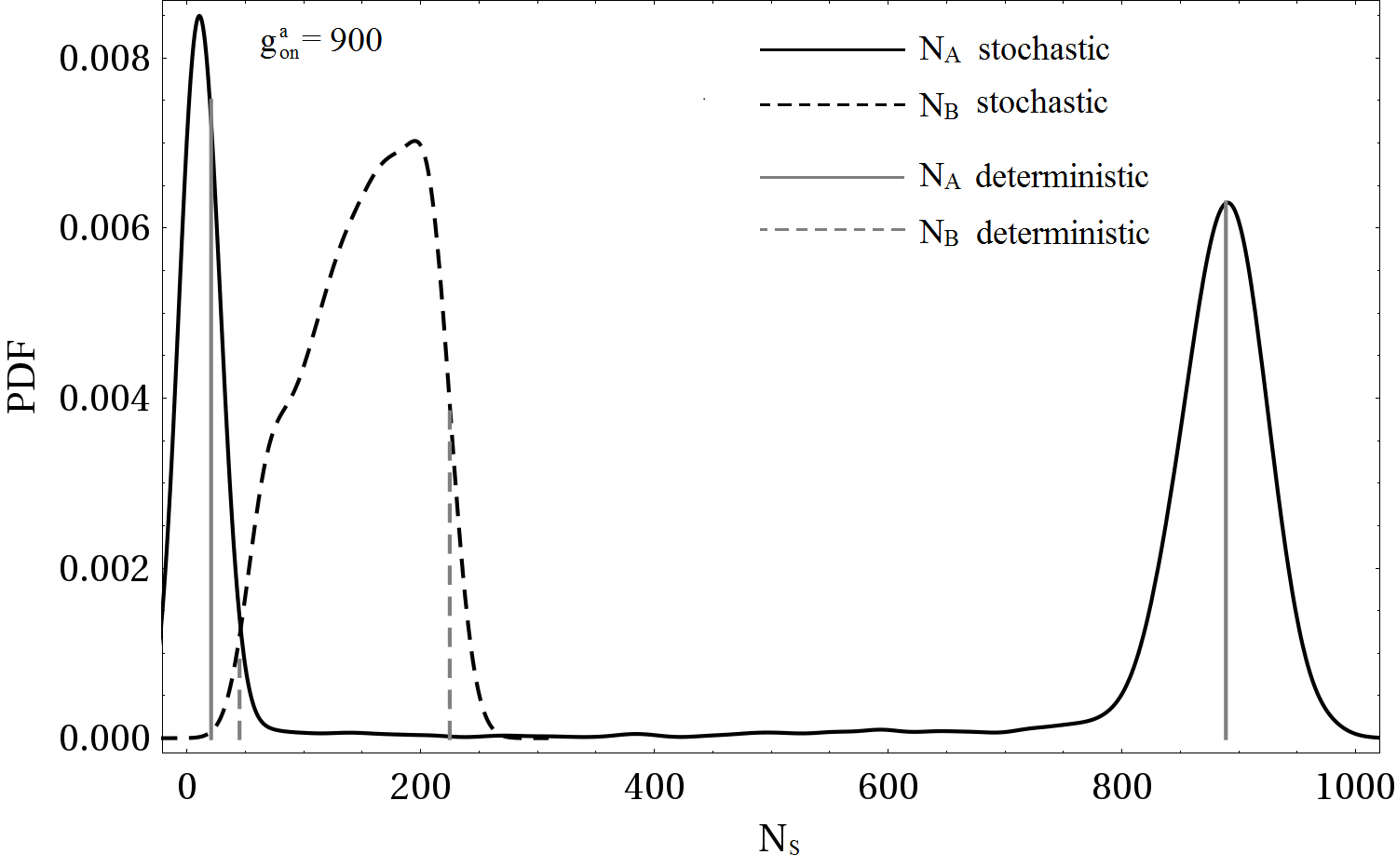}
    \caption{Same as Fig.~\ref{fig11}, but for slow genes, that is $h^a_{AA}=h^a_{BB}=0.0001$, $h^b_A=0.01$, $f^a_{AA}=f^a_{BB}=f^b_A=1.0$. The two vertical lines in each species in the PDFs indicate the prediction of two stable fixed points. This is in agreement with the stochastic results only for $g^a_{\emph{on}}=900$ (third row). For $g^a_{\emph{on}}=100$ (first row) we see the stochastic pendant of a single fixed point only in disagreement with the prediction from the deterministic model. The interpretation of the phase portrait in the second row remains ambiguous: the second local maximum in the PDF seems to be a precursor of the second fixed point, and the interpretation of the phase portrait in terms of a limit cycle is less likely due to the absence of a white area which should not be visited by the Gillespie trajectory in case of a limit cycle. For further explanations we refer to the main text.}\label{fig14}
  \end{centering}
\end{figure}

While for fast (ultra-slow) genes, gene switches are fast (slow) with respect to both proteins, for slow genes protein $A$ sees slow switches between the $\emph{on}-\emph{bare}$ and $\emph{off}-\emph{bare}$ average values, so that it can follow these different states of gene $a$, while protein $B$ is intrinsically so slow that it still sees averages over all states of the genes. Therefore inserting the separate average values of $\langle N_A\rangle$ in Eqs.~(\ref{eq24b}),(\ref{eq25b}) fails as an effective description. The phase portraits of $N_A,N_B$ for slow genes in the left column of Fig.~\ref{fig14} shows the pendant of one fixed point of Eq.~(\ref{eq25a}) (upper row), while the fixed point of Eq.~(\ref{eq24a}) is not visible (two fixed points each in the middle and lower row of the figure), where the left cloud corresponds to the solution of Eq.~(\ref{eq25a}) and the right cloud to Eq.~(\ref{eq24a}). In particular we interpret the clouds of events in the second row also as a stochastic switching between two fixed points rather than a noisy version of limit cycles, since in contrast to the corresponding phase portrait in Fig.~\ref{fig11} there is no empty space between large and small $N_A$ values, the events jump from the left-to the right side rather than performing full cycles, in agreement with the time series of $N_A$ in Fig.~\ref{fig13}. The vertical lines in Fig.~\ref{fig14} indicate the two stable fixed points of Eqs.~(\ref{eq24a}),(\ref{eq25a}) in all three cases. The second fixed point of Eq.~(\ref{eq24a}) is not visible for $g^a_{\emph{on}}=100$ (first row), but it is seen for $g^a_{\emph{on}}=300$ and $g^a_{\emph{on}}=900$, where both fixed point locations fit well the maxima of the PDFs for $N_A$, though not for $N_B$. The failure is due to the inherent difficulty that protein $B$ sees averages over all $b$-states, leading to values of $N_B$ which are not solutions of Eqs.~(\ref{eq24b}),(\ref{eq25b}) (derived under the condition on protein $A$ to see $N_A$ states as two distinct averages over ($\emph{on}-\emph{bare}$) and $\emph{off}-\emph{bare}$) states). $B$ is still too inert to follow the two distinct $\langle N_A\rangle$-values.

\subsection{Ultra-slow genes}\label{secIII3}
Independently of the possibility to realize this limit in natural or synthetic genetic circuits, it is of interest from the dynamical point of view what the effective coarse-grained description in the deterministic limit amounts to in case of ultra-slow genes. This limit refers to a situation, in which the binding/unbinding rates of transcription factors are of the order of the slow protein $B$. So the time which genes $a$ and $b$
spend in one of their possible states is long as compared to $1/\delta^A$, the lifetime of protein $A$.
\begin{figure}
  \begin{centering}
    \includegraphics[width=9cm]{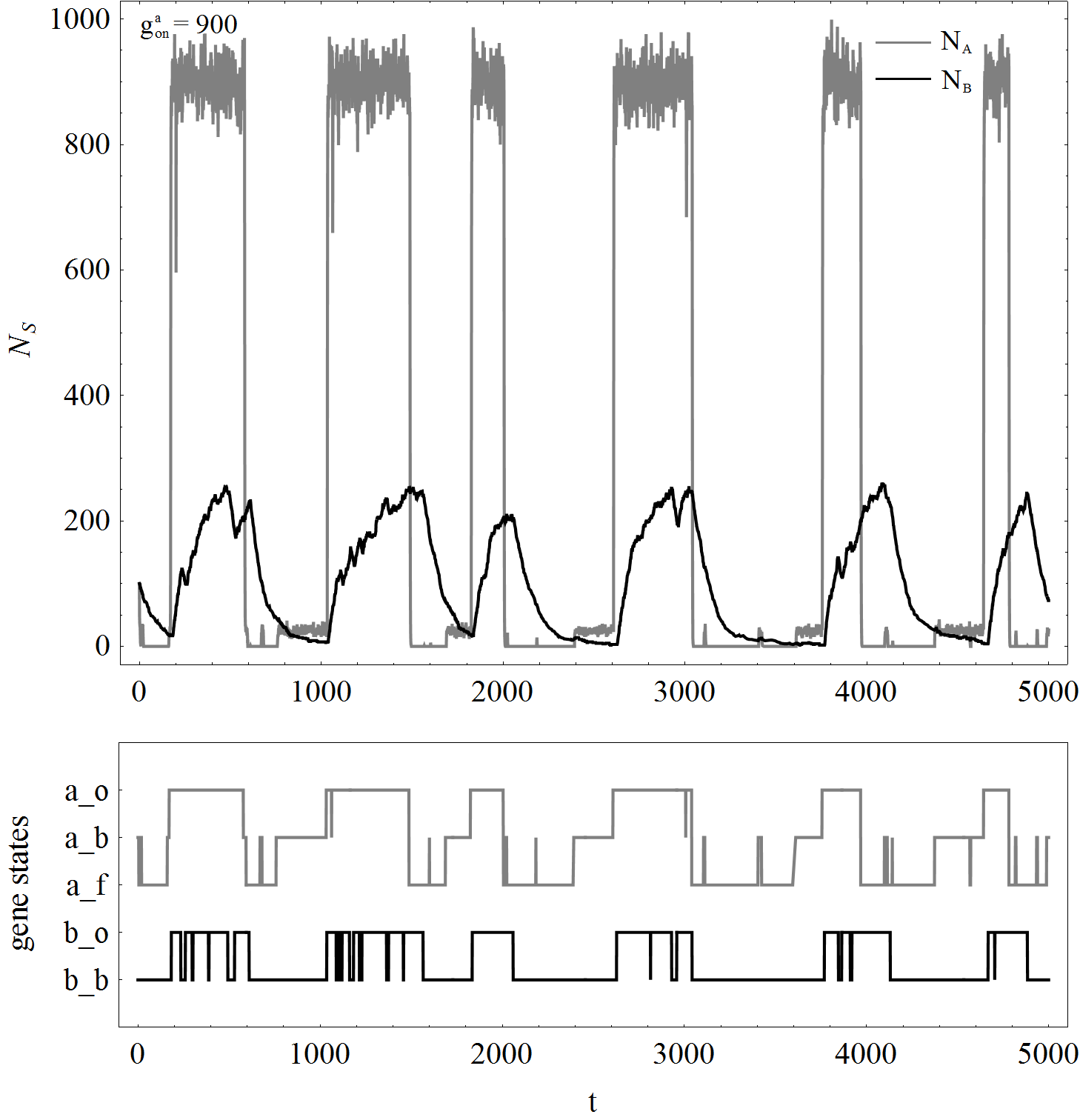}
    \par\end{centering}
  \caption{Same as Fig.~\ref{fig10}, but for ultra-slow genes, that is $h^a_{AA}=h^a_{BB}=0.000005$, $h^b_A=0.0001$, $f^a_{AA}=f^a_{BB}=f^b_A=0.01$. Protein $A$ sees gene $a$ in three different states, while there are periods during which protein $B$ still sees averages over $\emph{on}$ and $\emph{bare}$- states, here visible in the first two blocks, where gene $b$ switches frequently between $\emph{on}$ and $\emph{bare}$. }\label{fig15}
\end{figure}
\noindent It then does no longer make sense to further sum Eqs.~(\ref{eq13})--(\ref{eq15}) over any states of gene $a$ and Eqs.~(\ref{eq16}),(\ref{eq17}) over any states of gene $b$, but we are left with these five equations, which can be summarized as
\begin{equation}\label{eq18a}
  \frac{d(\langle N_S\rangle s_i)}{dt}\;=\;s_i\frac{d\langle N_S\rangle}{dt}\;+\;\langle N_S\rangle\frac{ds_i}{dt}
\end{equation}
with $S=A,B$, and $s_i=a_{\emph{on}},a_{\emph{bare}}, a_{\emph{off}}$ for $S=A$ and $b_{\emph{on}}, b_{\emph{bare}}$ for $S=B$, respectively.
If we insert for $ds_i/dt$ Eq.~(\ref{eq18}) and solve (\ref{eq18a}) for $d\langle N_S\rangle/dt$, we arrive at the following set of uncoupled differential equations for $\langle N_S\rangle/N_0=:\Phi_S$
\begin{eqnarray}
  \frac{d\Phi_A}{dt}&=&g^a_{i}\;-\;\delta^A\Phi_A\;, \;\;i=\emph{on},\;\emph{bare},\;\emph{off}\nonumber\\
  \frac{d\Phi_B}{dt}&=&g^b_{i}\;-\;\delta^A\Phi_A\;, \;\;i=\emph{on},\;\emph{bare}
\end{eqnarray}
with solutions that for $t\rightarrow\infty$ exponentially decay to the fixed points  $g^s_i/\delta^S$ with $S=A,B, s=a,b$, $i=\emph{on},\emph{bare},\emph{off}$ for $s=a$, and $i=\emph{on}, \emph{bare}$ for $s=b$, leading to six fixed points.

\subsubsection{Gillespie simulations for ultra-slow genes}\label{secIII31}
In the stochastic realization of this limit of ultra-slow genes we expect the system to switch between three possible states with respect to $N_A$ and two with respect to $N_B$, so between six fixed points in the deterministic limit. The former oscillations in the limit cycle regime are clearly gone. For $N_A$ we see both in the phase portraits and in the probability density functions remnants of three distinct fixed-point values of $N_A$, while the remnants of two possible fixed-point values of $N_B$ are only vaguely visible as two maxima in the probability distribution. Obviously the ultra-slow genes are still not slow enough to allow protein $B$ to adjust to the different states of gene $b$.

\begin{figure}
  \begin{centering}
    \includegraphics[width=6.5cm,height=4.5cm]{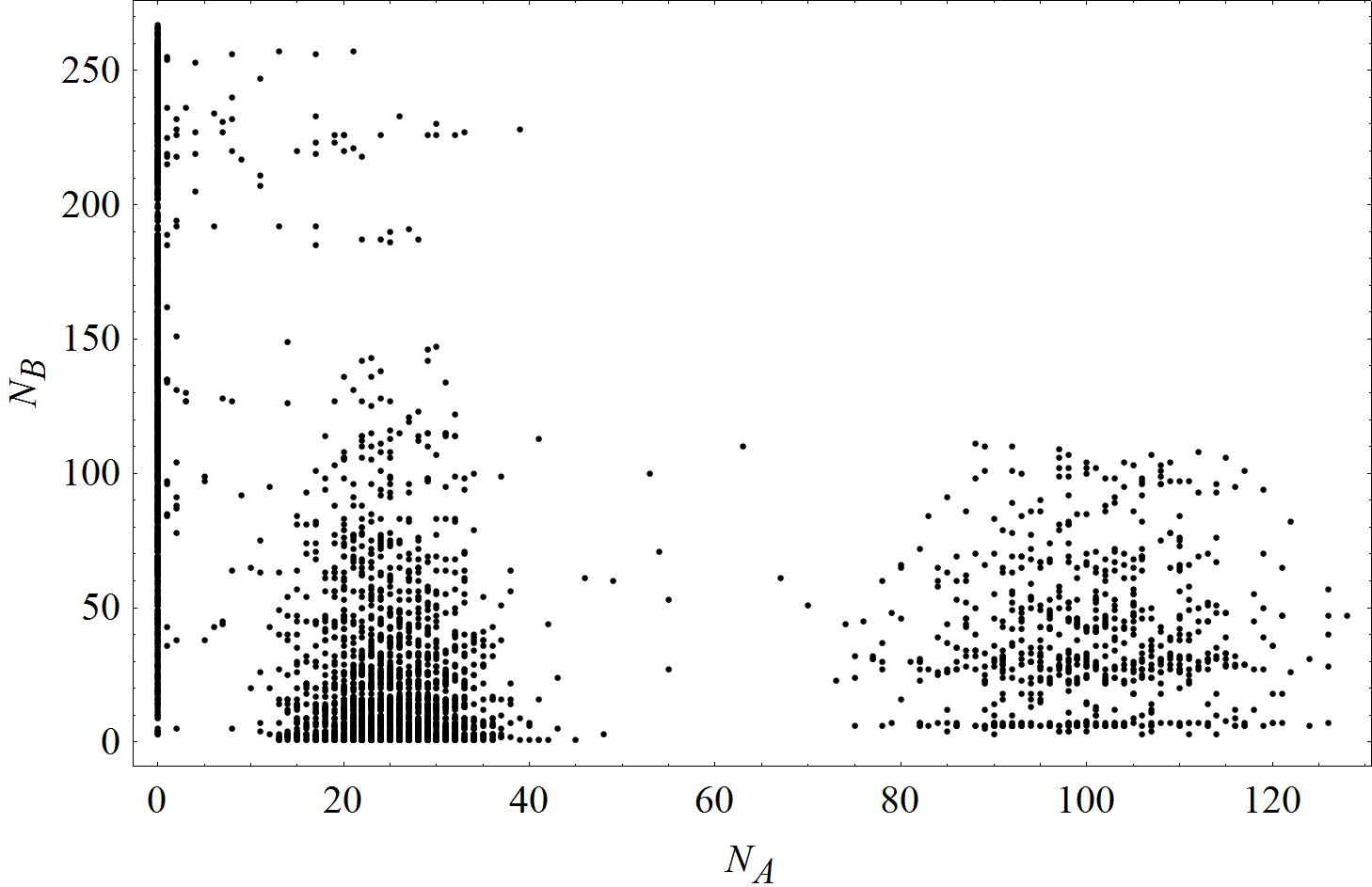}$\;$$\;$
    \includegraphics[width=6.5cm,height=4.5cm]{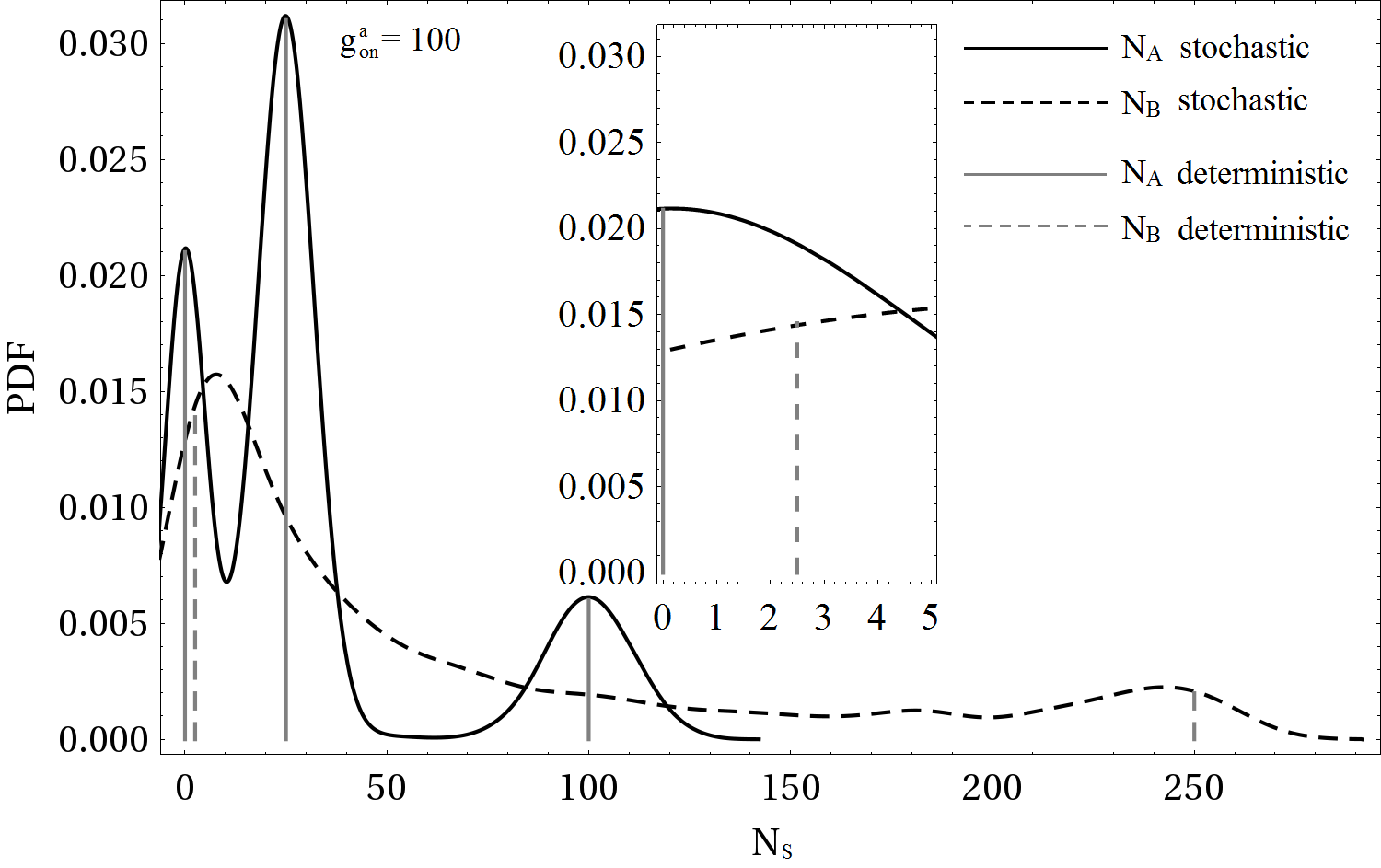}
    \par\end{centering}

  \begin{centering}
    \includegraphics[width=6.5cm,height=4.5cm]{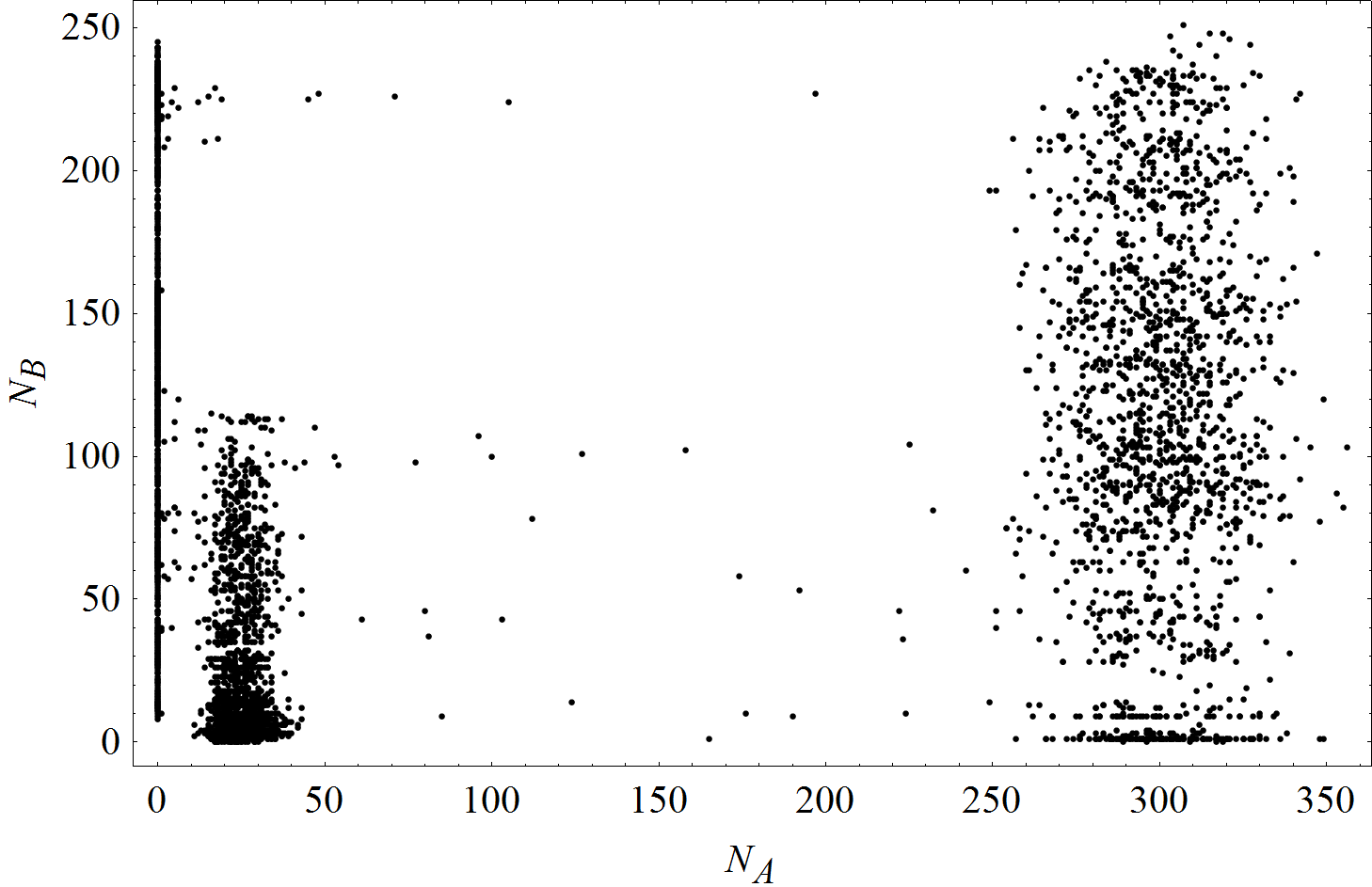}$\;$$\;$
    \includegraphics[width=6.5cm,height=4.5cm]{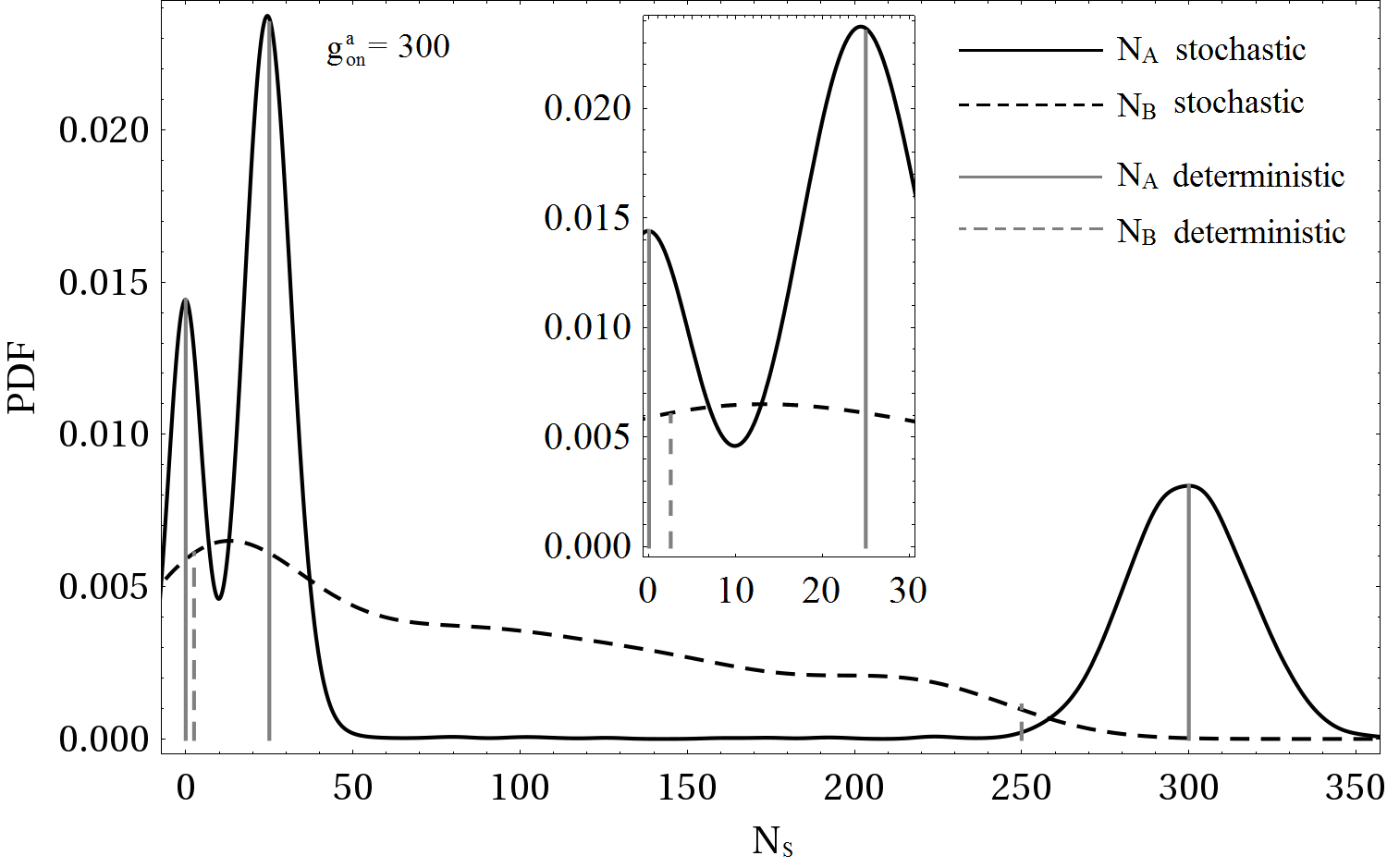}
    \par\end{centering}

  \begin{centering}
    \includegraphics[width=6.5cm,height=4.5cm]{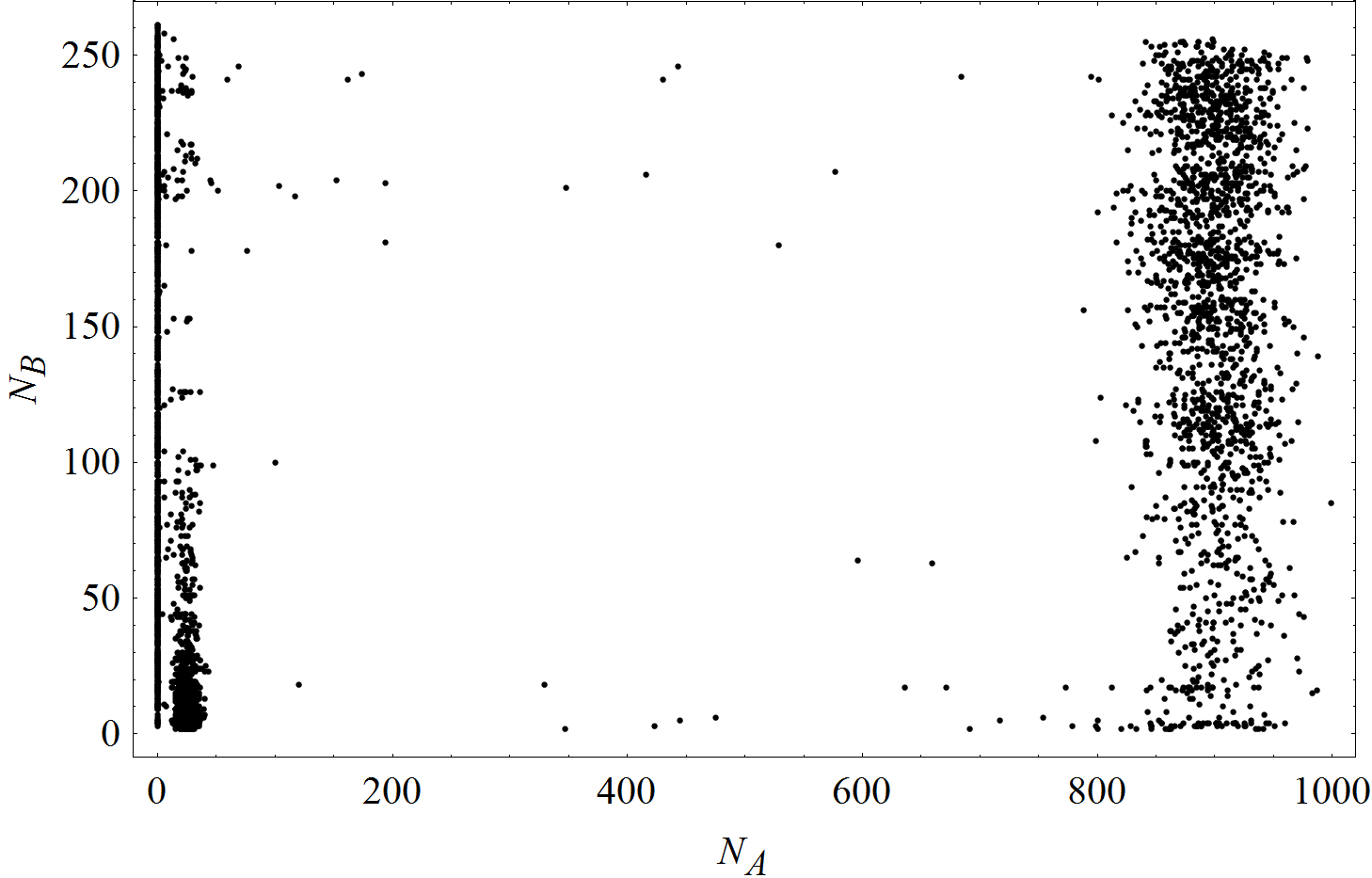}$\;$$\;$
    \includegraphics[width=6.5cm,height=4.5cm]{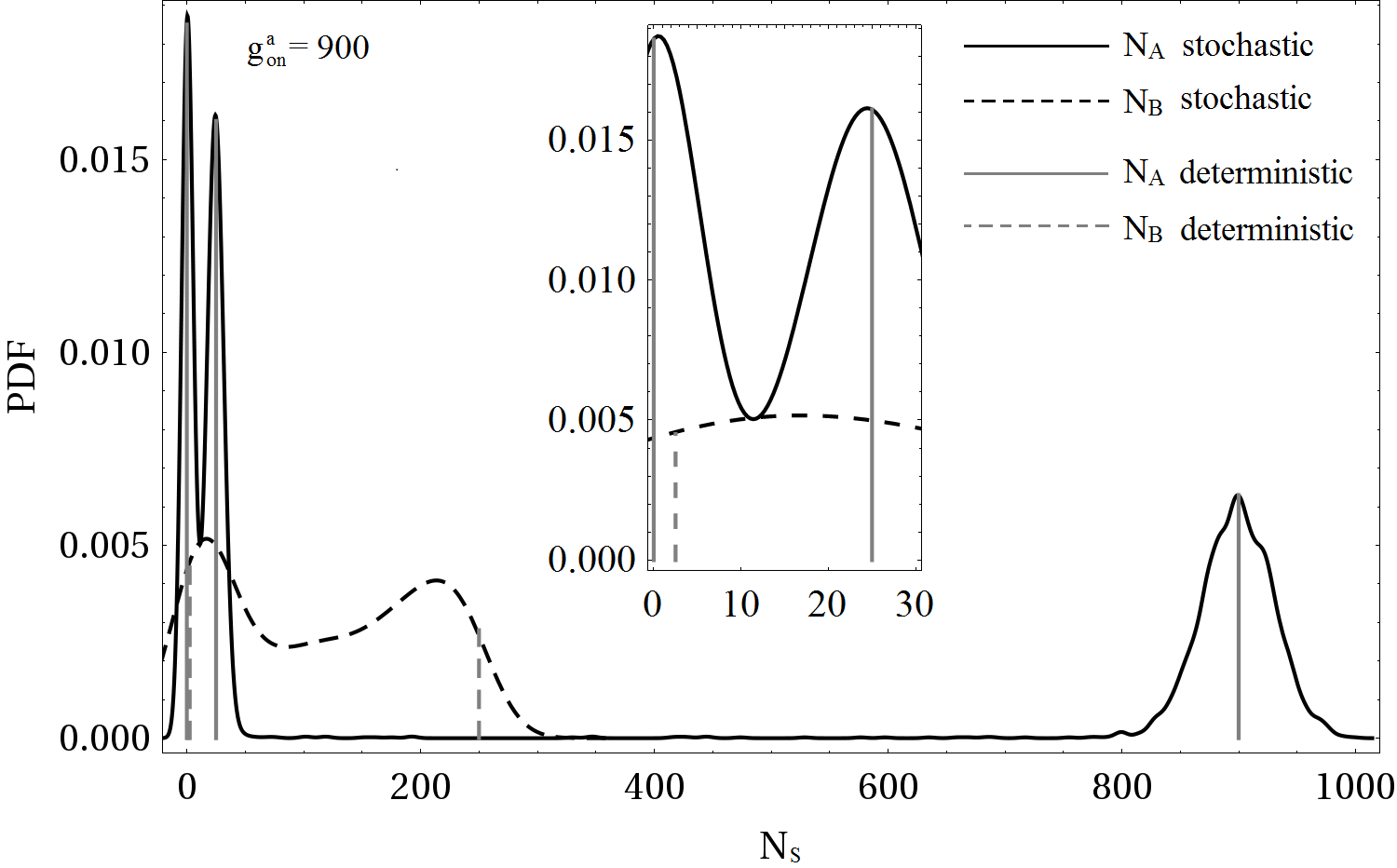}
    \caption{Same as Fig.~\ref{fig11}, but for ultra-slow genes, that is that is $h^a_{AA}=h^a_{BB}=0.000005$, $h^b_A=0.0001$, $f^a_{AA}=f^a_{BB}=f^b_A=0.01$. In all three rows, left column, we see remnants of three fixed points in the deterministic limit with respect to values of $N_A$, while the values of $N_B$ are broadly spread, since $B$ is too slow to follow the different states of gene $b$. The vertical lines from the deterministic prediction of the fixed point values (right column) match well the maxima of the PDFs for $N_A$ and only roughly for $N_B$. The insets zoom into the $(N_A,N_B)$ values of the two fixed points for lower  $N_A$-values.}\label{fig16}
  \end{centering}
\end{figure}

\section{Summary and conclusions}\label{secIV}

The motif of the self-activating species which activates its own repressor is found in many realizations in
biological systems. From the physics' point of view one would like to identify universal features of the dynamical behavior. Certainly regimes of excitable and oscillatory behavior are common features found over a wide range of parameters. Also in our current realization we have recovered three regimes of excitable, oscillatory and excitable behavior as function of one bifurcation parameter, which we have chosen as $g^a_{\emph{on}}$, the production rate of protein $A$ if gene $a$ is in the on-state. However, as we have pointed out, these three distinct regimes are only obtained in our realization if the binding and unbinding rates of genes are fast as compared to the other inherent time scales, here the decay rates of the fast $(A)$ and the slow $(B)$ proteins. For this case we derived deterministic equations equivalent to the former ones of \cite{pablohmo}. As soon as the binding/unbinding rates are no longer small, but of the same order as the decay time of either proteins, the averaging procedure for deriving a deterministic limit has to be changed; the proteins see no longer average values of the gene states, but tend to follow the distinct states, unless their own production is too slow to reach the appropriate state ``in time". These features were demonstrated by our Gillespie simulations of the six master equations. For the ultra-slow genes they were also visible in the derived coarse-grained description in terms of five deterministic equations. For the intermediate case of binding rates of the order of the decay of the fast protein, the intrinsic difficulty to derive deterministic equations lies in the fact that one and the same binding and unbinding events are differently seen from the proteins: protein $A$ sees already distinct events and follows the according gene state, protein $B$ still sees an average over the gene states. In both cases, slow and ultra-slow genes, the intermediate regime of stable regular oscillations is absent.

\noindent One should keep in mind that we have to deal with systems of nonlinear dynamics. Therefore one should be aware that apparently minor changes such as the value of the Hill coefficient may have pronounced effects. Even if the qualitative picture after such a minor change remains the same, the quantitative features like the extension of the limit-cycle regime can drastically change, as we have seen. Such a change can finally decide on the relevance of the model for the real biological system. For real systems stable oscillations would probably not be observed if they were restricted to a tiny interval of the bifurcation parameter.

\noindent Although we have chosen our parameters independently of their possible realization in concrete genetic circuits, the conclusion from our analysis is generic and applies also to real biological systems: It is not only the gross features of the topology of the motifs and the couplings that determine the dynamical performance. If different time scales are inherent, the gross bifurcation patterns may depend on their ratios.

\section*{Acknowledgment} We would like to thank Ashok Garai (UC San Diego) for his participation in the beginning of this work. Two of us (DL and HMO) would also like to thank Michael Zaks (Humboldt University Berlin) for stimulating discussions. Financial support from the DFG ((DL, grant no. ME-1332/17-1), (HN, grant no. JA 483/27-1)) is gratefully acknowledged.

\section*{Appendix: Zoom into the bifurcations of the deterministic equations for fast genes}\label{secIII13}
\begin{figure}
  \begin{centering}
    \includegraphics[width=10cm]{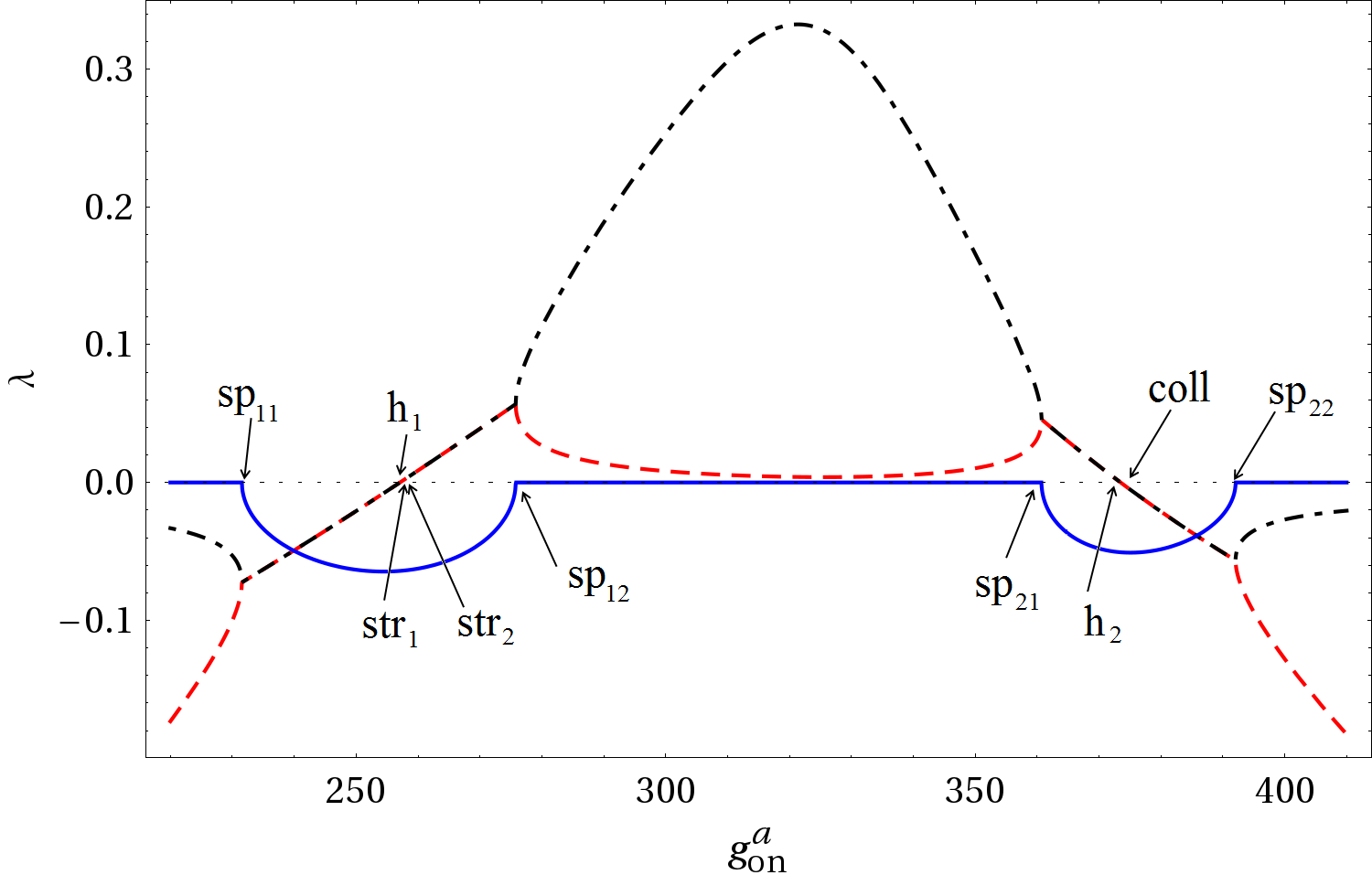}
    \par\end{centering}
  \caption{Real (black and red (gray)) and imaginary (blue) part of the eigenvalues of the Jacobian as function of the bifurcation parameter $g^a_{\emph{on}}$. To keep the figure clear we present only the negative imaginary part. For the meaning of the various indicated special values of $g^a_{\emph{on}}$ we refer to the main text.}\label{fig3}
\end{figure}

The appendix is devoted to a detailed bifurcation analysis of Eqs.~(\ref{eqdetfastgenes1}), (\ref{eqdetfastgenes}).
In Fig.~\ref{fig3} we plot the real-and imaginary part of the two eigenvalues of the Jacobian $\partial \dot{\Phi}_S/\partial \Phi_{S^\prime}, S,S^\prime\in\{A,B\}$, evaluated at the respective fixed points of these equations. We display only the negative imaginary part to keep the figure clear.  In the following we describe the stationary states as a function of increasing parameter $g^a_{\emph{on}}$. The labels assigned to the specific values of $g^a_{\emph{on}}$ shall remind to the change in the dynamical performance occurring at these values (so that spirals are created at ``sp", Hopf bifurcations at ``h", two limit cycles collide at ``coll" and ``str" marks the boundaries of the tiny ``strange" interval in which two limit cycles coexist).

\begin{itemize}
\item
  $g_{\emph{on}}^a<sp_{11}=231.5492$: one stable node. \\
  \noindent In this interval we observe one fixed point in the form of a stable node.
\item
  $sp_{11}=231.5492\leq g^a_{\emph{on}}<h_1=257.0829$: one stable spiral. \\
  \noindent
  At $sp_{11}=231.5492$ the two real eigenvalues become complex conjugate of each other, and the stable node turns into a stable spiral, shown in Fig.~\ref{fig4}.

  \begin{figure}
    \begin{centering}
      \includegraphics[width=10cm]{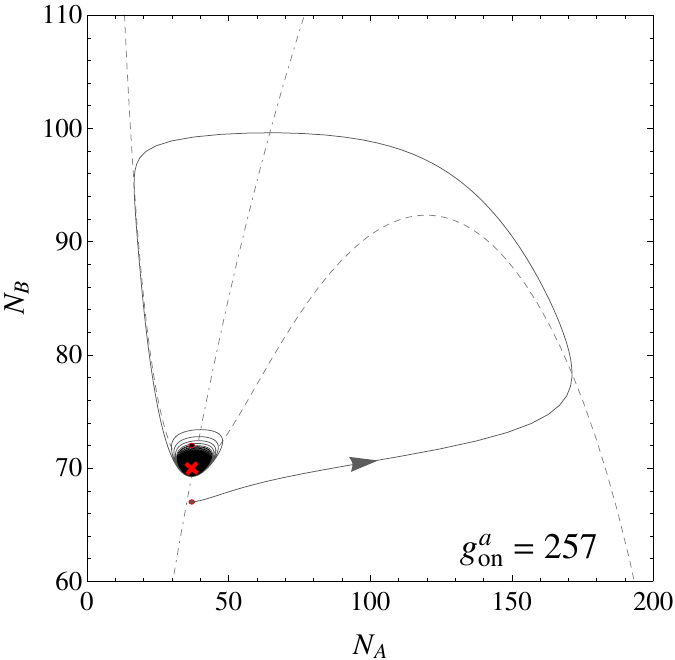}
      \par\end{centering}
    \caption{Stable spiral. Starting in the vicinity of the fixed point (red (gray) cross) either from above or from below (both starting points indicated by red dots), the trajectory spirals into the fixed point, either directly, when starting from above, or after a long excursion in phase space, starting from below. The other parameters are chosen as $g_{\emph{bare}}^{a}= 25$, $g_{\emph{off}}^{a}= 0$, $g_{\emph{on}}^{b} = 2.5$, $g_{\emph{bare}}^{b} = 0.025$, $h_{AA}^{a}$ = $h_{BB}^{a} = 0.01$, $h_{A}^{b} = 1$, $f_{AA}^{a}=$ $f_{BB}^{a}=$ $f_{A}^{b}$ $= 100$, $\delta^{A}= 1$, $\delta^{B} = 0.01$. Gray dashed and dashed-dotted lines show the nullclines of $N_A$ and $N_B$.}\label{fig4}
  \end{figure}

\item
  $h_1=257.0829\leq g^a_{\emph{on}}<str_1=257.12233$: one unstable spiral and one small stable limit cycle.\\
  \noindent At the Hopf bifurcation point $h_1=257.0829$ the real part of the eigenvalues crosses the abscissa and becomes positive, so that the stable spiral becomes unstable. From that value on, as long as $g^a_{\emph{on}}<str_1$, there is a small stable limit cycle with a radius growing with increasing $g^a_{\emph{on}}$, surrounding the unstable spiral. Trajectories spiral out of the fixed point towards the limit cycle and spiral extremely slowly when approaching the limit cycle, cf. Fig.~\ref{fig5}. The first Lyapunov coefficient $l_1(h_1)=-3.751804\cdot 10^{-6}$, where the negative sign of $l_1$ shows that the Hopf bifurcation is supercritical.

  \begin{figure}
    \begin{centering}
      \includegraphics[width=10cm]{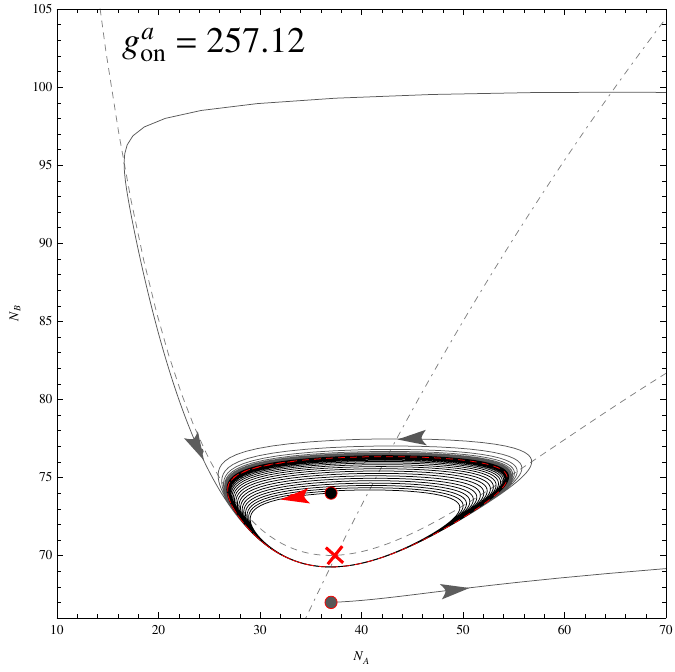}
      \par\end{centering}
    \caption{Unstable spiral and small stable limit cycle. Starting now in the vicinity of the fixed point (red (gray) cross) from above or below, the trajectory ends up in a small stable limit cycle indicated as red dashed line. Parameters other than $g^a_{\emph{on}}$ are chosen as in Fig.~\ref{fig4}.}\label{fig5}
  \end{figure}

  Without a further zoom into the bifurcation parameter, one would conclude from the observation of a large stable limit cycle for $g^a_{\emph{on}}>str_2=257.12253$ (see below) that the small stable limit cycle for $g^a_{\emph{on}}\geq h_1$ has continuously grown to this large size. What happens instead is described next.
\item
  $str_1=257.12233\leq g^a_{\emph{on}}<str_2=257.12253$: two coexisting stable limit cycles and one unstable spiral.\\
  \noindent Figure~\ref{fig6} shows a snapshot in bifurcation parameter space, just after the second limit cycle is born. The two black stripes in the inset consists of two limit cycles, a smaller and a larger one, and the white space in between is filled by the trajectories, independently of the initial conditions,  as long as this space is sufficiently small.

  \begin{figure}
    \begin{centering}
      \includegraphics[width=10cm]{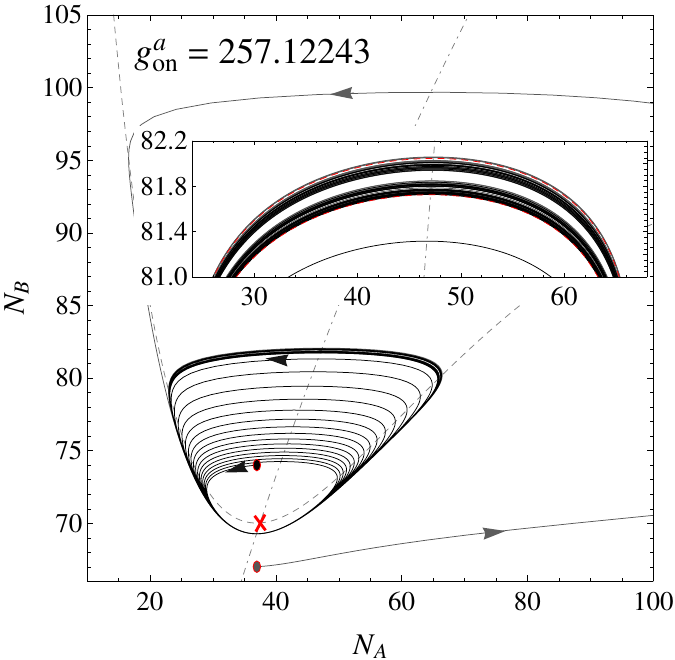}
      \par\end{centering}
    \caption{An instant in bifurcation parameter space, shortly after a second stable limit cycle is born. The coexistence of two stable limit cycles is visible as two black bands in the inset. Parameters other than $g^a_{\emph{on}}$ are chosen as in Fig.~\ref{fig4}.}\label{fig6}
  \end{figure}

  The second limit cycle pops up just after the small stable limit cycle from Fig.~\ref{fig5} has grown until $g^a_{\emph{on}}=str_1$. The phase space between the smaller and the larger limit cycle acts like an attractor as it is seen in Fig.~\ref{fig7}: trajectories, starting from initial conditions in this intermediate space (two initial conditions are indicated by the gray and black dot) follow limit cycles in the area between the small and the large limit cycle, indicated as gray dashed lines in the lower right part of the figure, see Fig.~\ref{fig7}. The zoom presented in the lower and upper right part shows two time snapshots up to time $T=2000$ and $T=10000$, respectively, where $T$ denotes the total sum over elementary time intervals.

  \begin{figure}
    \begin{centering}
      \includegraphics[width=13cm]{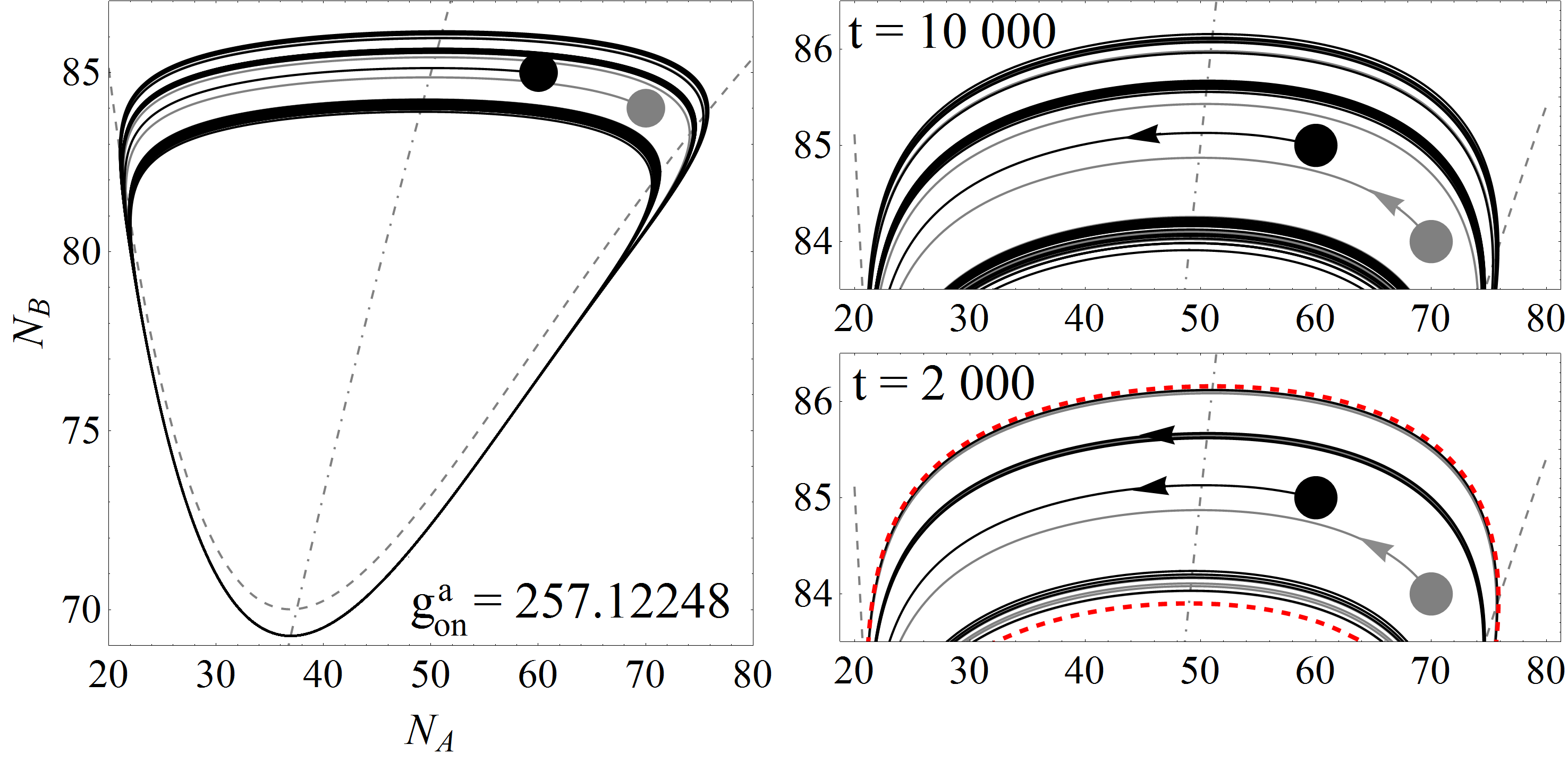}
      \par\end{centering}

    \centering{}
    \caption{
      Two coexisting limit cycles for $g^{a}_{\emph{on}}$ = 257.12248 and other parameters chosen as in Fig.~\ref{fig4}.
      % Other parameters: $g_{\emph{bare}}^{a}$ = 25, $g_{\emph{off}}^{a}$ = 0, $g_{\emph{on}}^{b}$ = 2.5, $g_{\emph{bare}}^{b}$ = 0.025,
      % $h_{AA}^{a}$ = $h_{BB}^{a}$ = 0.01, $h_{A}^{b}$ = 1, $f_{AA}^{a}$
      % = $f_{BB}^{a}$ = $f_{A}^{b}$ = 100, $\delta^{A}$ = 1, $\delta^{B}$ = 0.01.
      % Black and gray curves are trajectories with different initial conditions marked with  black and gray dots. %Arrows on the
      % indicate direction of the trajectory.
      Trajectories starting between the lower and upper limit cycle remain in this area bounded by the red (gray) dashed lines in the lower right part of the figure.
    }\label{fig7}
  \end{figure}

  The area between the limit cycles rapidly grows (Fig.~\ref{fig8} left), until only one large stable limit cycle remains (Fig.~\ref{fig8} right).

  \begin{figure}
    \begin{centering}
      \includegraphics[width=6.5cm]{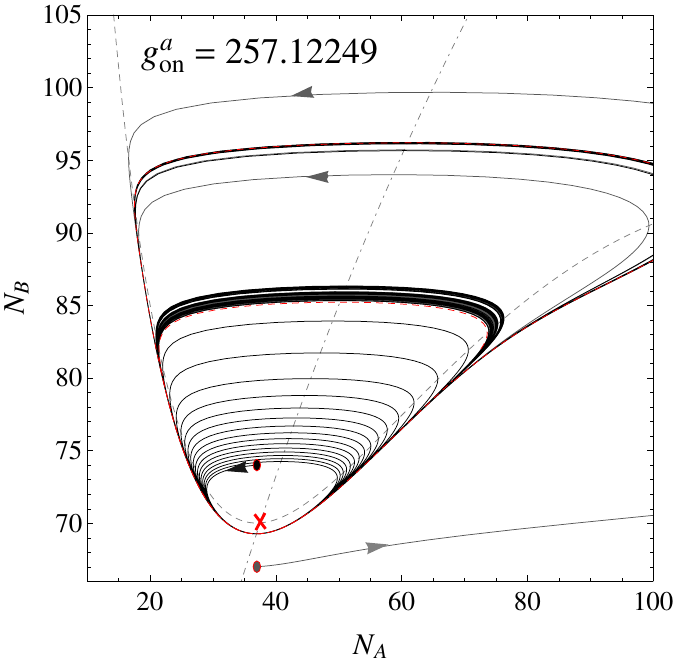}
      \includegraphics[width=6.5cm]{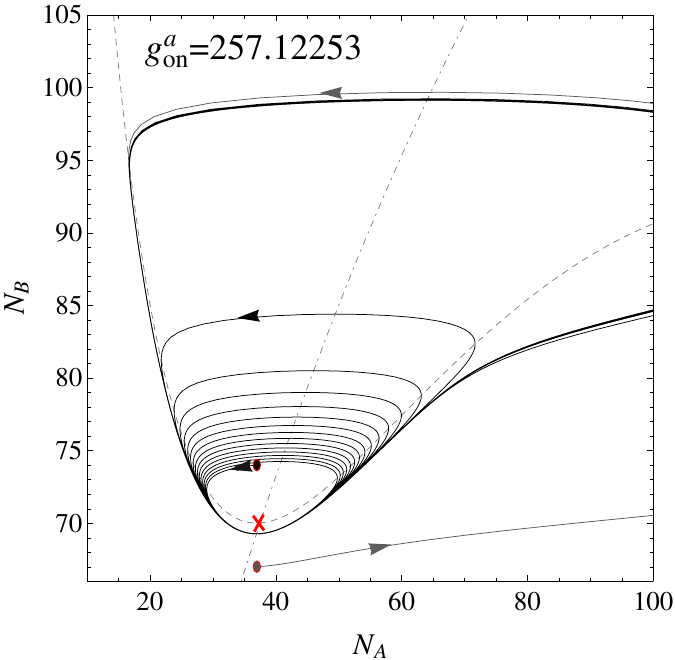}
      \par\end{centering}
    \caption{Left: Independently of the starting point, the trajectory will first hit the closer one of both limit cycles, then it will not stay exactly on this cycle, but continue in its vicinity. Right: From this value of $g^a_{\emph{on}}$ on, only one large limit cycle survives.}\label{fig8}
  \end{figure}

\item
  $str_2=257.12253\leq g^a_{\emph{on}}<sp_{12}=257.7795$: one large stable limit cycle and one unstable spiral. \\
  \noindent From $g^a_{\emph{on}}=str_2$ on only one large stable limit cycle is left along with the unstable fixed point which stays a spiral until $g^a_{\emph{on}}=sp_{12}$.

  % \begin{figure}[H]
  %   \begin{centering}
  %     \includegraphics[width=10cm]{LC25712253str.pdf}
  %     \par\end{centering}
  %   \caption{Real}\label{fig9}
  % \end{figure}

\item
  $sp_{12}=257.7795\leq g^a_{\emph{on}}<sp_{21}=360.71422$: one large stable limit cycle and one unstable node.\\
  \noindent At $g_{\emph{on}}^a=sp_{12}$, the imaginary part of the complex conjugate eigenvalues become zero, leaving two real different positive eigenvalues, so that the fixed point inside the large stable limit cycle has turned into an unstable node. This situation holds over a large interval in $g^a_{\emph{on}}$, until the unstable node turns back into an unstable spiral, not displayed.
\item
  $sp_{21}=360.71422\leq g^a_{\emph{on}}<h_2=373.4836$: one large stable limit cycle and one unstable spiral. \\

  % \begin{figure}[H]
  %   \begin{centering}
  %     \includegraphics[width=10cm]{LC3734.pdf}
  %     \par\end{centering}
  %   \caption{Real}\label{fig11}
  % \end{figure}

  \noindent The return to an unstable spiral happens at $g^a_{\emph{on}}=sp_{21}$, where the eigenvalues become again complex conjugate. The spiral remains unstable until the second Hopf bifurcation happens.
\item
  $h_2=373.4836\leq g^a_{\emph{on}}<coll=373.90582$: one stable spiral, one small unstable limit cycle and one large stable limit cycle.\\
  \noindent The second Hopf bifurcation happens at $g^a_{\emph{on}}=h_2$, where the real part of the complex conjugate eigenvalues crosses the abscissa, the formerly unstable spiral changes into a stable one. Along with that a small unstable limit cycle is born, while the large stable limit cycle is still ``alive". So we have a stable spiral, surrounded by a small unstable limit cycle, surrounded by the large stable limit cycle (not visible in Fig.~\ref{fig9}), see Fig.~\ref{fig9}.

  \begin{figure}
    \begin{centering}
      \includegraphics[width=10cm]{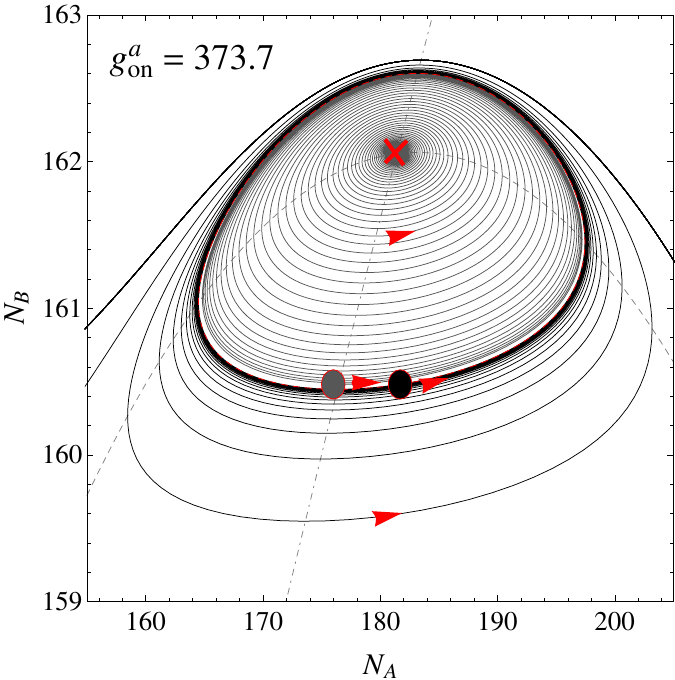}
      \par\end{centering}
    \caption{Two starting points inside and outside the unstable limit cycle lead to a spiral into the fixed point and to the large limit cycle (not displayed), respectively. Parameters other than $g^a_{\emph{on}}$ are chosen as in Fig.~\ref{fig4}.}\label{fig9}
  \end{figure}

  \enlargethispage{\baselineskip}

  This coexistence holds only for a tiny interval, as long as $g^a_{\emph{on}}<coll$. The reason for the small size of this interval is that the fixed point and the small unstable limit cycle are located close to the trajectory of the large stable limit cycle, so that a tiny increase of $g^a_{\emph{on}}$ leads to a growth of the radius of the small unstable cycle that is sufficient for the small cycle to collide with the large cycle and this way to induce its cancelation. The collision of the two limit cycles amounts to a saddle-node bifurcation of two periodic orbits with the result that the periodic orbits disappear. This suggests the vicinity of a Bautin bifurcation \cite{bautin}; the Bautin bifurcation is characterized by two bifurcation conditions: (i) $Re \lambda_{1,2}=0$ and (ii) the critical first Lyapunov coefficient $l_1(\alpha_{\text{crit}})=0$. We have in our interval $h_2\leq g^a_{\emph{on}}<coll$ that $Re \lambda_{1,2}=0$ and $l_1(h_2)=9.5\cdot 10^{-6}$ at the Hopf bifurcation point.
\item
  $coll=373.90582\leq g^a_{\emph{on}}<sp_{22}=392.0632$: one stable spiral, no limit cycle anymore.\\
  \noindent After the stable and unstable limit cycles disappear  at $g^a_{\emph{on}}=coll$, there remains a stable spiral as long as $g^a_{\emph{on}}<sp_{22}$.
\item
  $sp_{22}=392.0632\leq g^a_{\emph{on}}$: one stable node.\\
  \noindent At $g^a_{\emph{on}}=sp_{22}$ the two complex conjugate eigenvalues  become real again with two different negative values, so that finally the stable spiral returns to a stable node and remains like that for increasing $g^a_{\emph{on}}$.
\end{itemize}
\enlargethispage{\baselineskip}
While the linear stability analysis  reveals the change from a stable spiral to an unstable spiral, and an unstable node back to an unstable spiral, to a stable spiral and a stable node, the numerical integration of the equations (\ref{eqdetfastgenes1}),(\ref{eqdetfastgenes}) shows subtleties in how the large stable limit cycle is born and destroyed. The creation happens discontinuously and in coexistence with a small stable limit cycle, the annihilation very likely via a Bautin bifurcation.

%%%%%%%%%%%%%%%%%%%%%%%%%%%%%%%%%%%%%%%%%%%%%%%%%%%%%%%%%%%%%%%%%%%%%%%%%%%%%%%%%%%%%%%%%%%%%%% 
%%%%%%%%%%%%%%%%%%%%%%%%%%%%%%%%%%%%%%%%%%%%%%%%%%%%%%%%%%%%%%%%%%%%%%%%%%%%%%%%%%%%%%%%%%%%%%%%% 

{}

\end{document}